# Domain Formation and Correlation Effects in Quenched Uniaxial Ferroelectrics: A Stochastic Model Perspective

Olga Y. Mazur[1,2,*], Yuri A. Genenko[2], Leonid I. Stefanovich[3]

[1] Institute of Mechatronics and Computer Engineering, Technical University of Liberec, Studentská 1402/2, 46117 Liberec 1, Czech Republic

[2] Institute of Materials Science, Technical University of Darmstadt, Otto-Berndt-Str, 3, 64287 Darmstadt, Germany

[3] Branch for Physics of Mining Processes of the M.S. Poliakov Institute of Geotechnical Mechanics of the National Academy of Sciences of Ukraine, Simferopolska 2a, 49600 , Dnipro, Ukraine

[*]**Author to whom correspondence should be addressed:** o.yu.mazur@gmail.com, +380509142572

**ABSTRACT.**

The stochastic analysis of the polarization domain structures, emerging after quenching from a paraelectric to a ferroelectric state, in terms of the polarization correlation functions and their Fourier transforms is a fast and effective tool of the materials structure characterization. In spite of a significant volume of experimental data accumulated over the last three decades for the model uniaxial ferroelectric triglycine sulfate, there were no theoretical tools to comprehend these data until now. This work summarizes the recent progress in understanding of the experiments by means of the original stochastic model of polarization structure formation based on the Landau-Ginzburg-Devonshire theory and the Gauss random field concept assuming the predominance of the quenched polarization disorder over the thermal fluctuations. The system of integrodifferential equations for correlation functions of random polarization and electric field turns out to be analytically solvable. The model provides explanations to a range of experimental results on the polarization formation kinetics including the time-dependent correlation lengths and correlation functions on the macroscopic spatial and time scales. Notably, it predicts the dependence of the ferroelectric coercive field on the initial disordered state characteristics, which can be controlled by quenching parameters like the initial temperature and the cooling rate, thus paving the way for tailoring the functional properties of the material.

## 1. INTRODUCTION

Versatile physical properties of ferroelectric materials find a wide application in photonics, nonlinear optics, acoustics, nanoelectronics and photovoltaics [1 – 11]. High dielectric constant and polarization switching are used in memory modules for non-volatile data storage and compact autonomous ultrahigh power density energy storage [2, 3]. The ultrafast polarization reversal (tens of nanoseconds) consuming small energy is a major advantage of ferroelectric materials for synaptic device application in neuromorphic computing [4, 5]. Pyroelectric effect is exploited in infrared (IR) sensors for detecting IR radiation emitted by remote objects [6, 7] and pyroelectric micro heat engines for energy harvesting [8]. Many ferroelectric materials exhibit the photovoltaic effect due to the asymmetric distribution of charges within a crystal structure. The polarization switching can facilitate the separation of photogenerated electron-hole pairs and direct charge carriers towards the respective electrodes, reducing the recombination losses and enhancing the efficiency of light absorption. High stability and robustness against degradation by temperature, radiation and humidity make ferroelectric components promising materials for future solar cell devices [9 – 11].

The decisive role that the domain structure plays in all ferroelectric properties in demand cannot be overestimated [12 – 20]. Different size, shape and orientations of domains, dynamics of domain walls and polarization switching provide variable dielectric constants, pyroelectric and piezoelectric coefficients, affecting the capacity, response time, hysteretic and transient properties [17, 18]. In particular, the maximum pyroelectric coefficient is observed in single-domain samples, but the nonhomogeneous structure with a lot of fast-moving domain walls can also enhance the pyroelectricity and improve the thermal sensitivity of the material [19]. Understanding and controlling the domain structure are thus essential for optimizing ferroelectric materials for diverse applications, ranging from memory devices and sensors to actuators and energy harvesting systems.

Polarization domain structure arises after quenching from a high-temperature paraelectric phase into a low-temperature ferroelectric one. This is a stochastic process which depends on material properties, prehistory of the sample, quenching conditions and external fields [21 – 29]. Such a variety of factors leads to inconsistencies in observed properties of the same materials and complicates the development of theoretical approaches predicting the domain dynamics. Valuable insights into the complex mechanisms of domain nucleation, growth and coarsening are provided by simulation methods, such as micro-mechanical and phase-field models [30-38], density-functional theory and effective Hamiltonian approaches [30,39], molecular dynamics (MD) and Monte-Carlo (MC) calculations [16, 40, 41]. Considering the influence of complex effects such as electrostriction, flexoelectric coupling and dipole-dipole interaction, these methods allow a deep insight into the emerging domain structures. The primary challenge here lies in the limited temporal scale (ranging from picoseconds to nanoseconds) and spatial resolution (from atomic to mesoscopic level), which complicates the application of simulations to real objects at macroscopic space and time scales.

Experimental observation of the kinetics of domain structure formation over macroscopic periods of time soon after quenching in the vicinity below $T_C$ has a number of technological limitations and was performed so far only for ferroelectrics of triglycine sulfate (TGS) group [42 – 56]. TGS is a uniaxial non-ferroelastic ferroelectric with a simple 180º domain structure, high dielectric permittivity and a single second-order phase transition that makes it the best model object both for theoretical and experimental studies. The first long-term observation of domain kinetics in TGS was performed by Tikhomirova *et al.* using the liquid-crystal method [42]. However, this work was mostly devoted to the investigation of domain wall dynamics under the influence of the sinusoidal electric field instead of the evolution after the quench. Further experiments by Nakatani [43] with the same observation method disclosed the detailed domain formation and growth in the vicinity of Curie point with $\Delta T = T_C - T = 0.39$ K. A lot of further measurements of the time evolution of the mean domain size were performed in the temperature range $0.03$ K $< \Delta T < 3$ K [47 – 56], i.e. just outside the range where the thermal fluctuations prevail ($\Delta T < 0.02$ K) [57], and showed the linear growth in time until it reaches 50 μm. The investigation by Tomita *et al.* was carried out far from the Curie point, $\Delta T = 25$ K [44] and provided the first characterization of domain correlations in the spirit of the Ising model for the non-conserved order parameter. The correlation length, longitudinal and transverse correlations were evaluated and the time evolution of the characteristic length was fitted by the power law $L(t) \sim (t - t_0)^y$, with the value $y = 0.218$. During thirty years this experiment was a basic approach for many further temporal investigations using various conditions and measuring techniques revealing the $y$ value to be from 0.086 to 0.3 [46 – 4951, 54]. *In situ* observations of domain wall dynamics in TGS during the phase transition have been done by the second-harmonic generation (SHG) microscopy allowing for the 3D domain structure imaging [53]. A rather detailed investigation of domain structure formation was conducted by Drozhdin *et al.* using the atomic force microscopy (AFM) for different TGS samples (pure, doped with alanine, irradiated) in a wide range of quenching temperatures [51, 55]. However, all these experimental results were still analyzed in the spirit of the Ising model as in the pioneering approach by Tomita *et al* [44].

A phenomenological stochastic theory for the applied field-driven ferroelectric domain reversal in TGS was elaborated by Ishibashi et al. [58] based on the Kolmogorov description of the crystal growth [59]. Within this approach, the frequency dependent polarization loops in TGS were evaluated by Hashimoto et al. [60], and polarization domain structure formation in lithium niobate ($LiNbO_3$) subject to a spatially modulated electric field was studied by Kalkum et al. [61]. All these theories did not account for the emerging electric depolarization fields and did not consider correlation properties of domain structures.

For the description of domain structure formation in ferroelectrics after quenching, the development of a stochastic model within the framework of the Landau-Ginzburg-Devonshire (LGD) theory seems suitable. The first stochastic analysis of domain growth based on the time-dependent LGD equation for the scalar order parameter was performed by Rao *et al.* [62] with the account of the thermodynamic fluctuations and the frozen disorder due to defects. The pair correlation function was calculated numerically and the diffusive regime of correlation length $L \sim t^{1/2}$ was revealed. An LGD-based stochastic approach for description of the multidomain ordering kinetics in ferroelectrics, solid solutions and thin films on solid substrates was proposed by Stefanovich *et al.* [63 – 65]. Another LGD-based theory of domain structure formation with a self-consistent account of the emerging depolarization field was developed by Darinskii et al. [66], however, the transient growth of the structure and its correlation properties were not studied. A thorough review of phase-ordering kinetics theories for the systems with not only usual scalar order parameter, but also vector and tensor fields, suitable for describing liquid crystals was given by Bray [67] but without account of the emerging electric depolarization fields.

Motivated by the diversity of experimental data for the domain growth in TGS crystals remaining unexplained [42 – 56], we advanced recently a stochastic theory of uniaxial ferroelectrics with a self-consistent account of interactions between the random polarization and electric fields [68 – 70]. This provided the formulation of the system of nonlinear integrodifferential equations for correlation functions of polarization and electric field. The problem turned out to be partially solvable analytically, allowing one to find the correlation coefficients in closed form. The full analysis requires numerical solution of the system of nonlinear differential equations for the polarization and its variance. It provides the next outcomes: all stages of the domain formation and growth were described; analytical expressions for the correlation length and the full set of correlation functions were derived; the methodological tool for prediction of the domain states in different quenching conditions and external effects was developed. This paper summarizes the analytical and numerical results from the exactly solvable isotropic stochastic model of uniaxial single crystal ferroelectrics in comparison with available experiments and demonstrates different possible transient scenarios of the domain structure formation with the focus on the polarization and electric field correlations relevant from both the fundamental and the practical points of view.

## 2. GENERAL DESCRIPTION OF THE MODEL

Many experiments on quenching TGS crystals were made on the arrangement consisting of a dielectric layer, in particular (few micrometers thick) liquid crystal placed for the visualization of the domain structure on the ferroelectric slab located between two metallic electrodes [43 – 48, 50, 53 – 55], that determines the choice of the geometry of the problem in this study (Fig. 1). The relation between dielectric and ferroelectric layers affects the rate and character of domain evolution and is specified by the parameter $a_z = h_d/(\varepsilon_d h_f + \varepsilon_b h_d)\varepsilon_0$, where $h_d$ and $h_f$ are thicknesses of the dielectric and ferroelectric layers, respectively: $\varepsilon_0$, $\varepsilon_d$ and $\varepsilon_b$ are the permittivity of vacuum, of the dielectric layer and the background permittivity of the ferroelectric, respectively (Fig. 1).

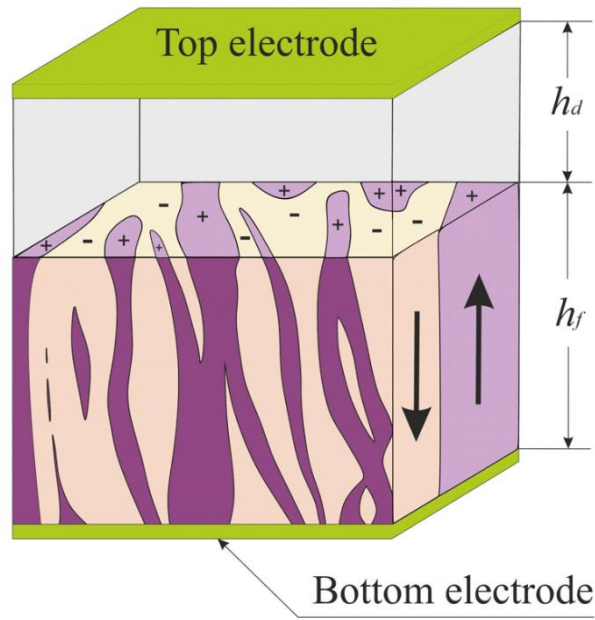

Figure 1. A ferroelectric crystal of thickness $h_f$ is placed between two electrodes and separated from the top by a dielectric layer of thickness $h_d$. The ferroelectric slab is infinite in the (*x, y*) plane and the polar axis is along the *z* direction in the Cartesian (*x, y, z*) frame.

We consider the evolution of the system from the initial state emerged after cooling from the paraelectric phase to the ferroelectric one at temperature $T < T_C$. The domain structure development occurs under a constant voltage applied to the electrodes as indicated in Fig. 1. The LGD energy functional for the uniaxial non-ferroelastic ferroelectric with a single polarization component along polar axis *z* undergoing a single second order phase transition can be represented to a first approximation in the form [68] neglecting the influence of the crystal anisotropy [71]:

$$\Phi = \Phi_0 + \int_{V_f} \left[\frac{1}{2}AP_z^2 + \frac{1}{4}BP_z^4 + \frac{1}{2}G(\nabla P_z)^2 - P_z E_z - \frac{\varepsilon_0 \varepsilon_b}{2}\boldsymbol{E}^2\right]dV - \int_{V_d} \frac{\varepsilon_0 \varepsilon_d}{2}\boldsymbol{E}^2 dV \qquad (1)$$

with the coefficient $A = \alpha_0(T - T_C)$, $\alpha_0 > 0$, $T < T_C$, which is the temperature of the second order paraelectric-ferroelectric phase transition, and the other temperature independent coefficients $B > 0$ and $G > 0$. $\boldsymbol{E}$ denotes the local electric field arising due to the applied voltage and the spatial variation of the polarization, while $V_f$ and $V_d$ denote the volumes of the ferroelectric plate and the dielectric layer, respectively. This form of the LGD functional significantly advances the developed stochastic theory with respect to the previous approaches [62, 63, 72] by the self-consistent account of the emerging depolarization field that makes possible observation of multidomain polarization states at finite external electric fields.

It is convenient for the following calculations to introduce dimensionless physical variables normalized to their natural characteristic magnitudes in the phase transition problem. Thus, we denote a dimensionless polarization $\pi = P_z/P_s$ normalized to the spontaneous equilibrium polarization $P_s = \sqrt{|A|/B}$, and a dimensionless electric field $\epsilon = \boldsymbol{E}/E_0$ with the value of $E_0 = P_s|A|$. All spatial coordinates are normalized to a characteristic length $\lambda = \sqrt{G/|A|}$ being the characteristic domain wall thickness. Since the initial state is a random one, all physical variables become random too and are considered in this model as Gauss random fields as suggested previously [63, 72]. Then the polarization can be represented as $\pi(\mathbf{r}, \tau) = \bar{\pi}(\tau) + \xi(\mathbf{r}, \tau)$ with the dimensionless mean polarization magnitude $\bar{\pi}(\tau) = \langle P_z \rangle / P_s$ depending on the dimensionless time $\tau$ and the stochastic polarization $\xi(\mathbf{r}, \tau)$, such that $\langle \xi(\mathbf{r}, \tau) \rangle = 0$. Here the sign $\langle ... \rangle$ denotes statistical averaging over all possible system realizations. Then a dimensionless local electric field in the chosen sample geometry reads

$$\epsilon(\mathbf{r},\tau) = \epsilon_a - \alpha_z \bar{\pi}(\tau)\hat{\mathbf{z}} - \boldsymbol{\nabla}\phi(\mathbf{r},\tau) \tag{2}$$

where $\epsilon_a = \varepsilon_d V/\big((\varepsilon_d h_f + \varepsilon_b h_d)E_0\big)$ is a uniform electric field in the ferroelectric induced by a voltage $V$ applied to the electrodes; $\alpha_z = a_z/|A|$ is the depolarization coefficient characterizing the mean depolarization field in the ferroelectric due to the average polarization $\bar{\pi}$; $\boldsymbol{\nabla}\phi(\mathbf{r},\tau)$ is the contribution of the stochastic electric depolarization field due to the stochastic electric potential $\phi(\mathbf{r},\tau)$, such that $\langle \boldsymbol{\nabla}\phi \rangle = 0$. We note here that the effect of thermal fluctuations can be neglected in a wide temperature range in comparison with that of the quenched disorder if the amplitude and the spatial scale of the latter are large enough [68], while the former dominates the system behavior in the close vicinity of the transition temperature, $T_C - T < 0.02$ K for TGS [57].

By variation of the energy functional (1) with respect to the polarization and the electric potential, respectively, a system of evolution equations can be derived [68],

$$\begin{cases} \dfrac{\partial \pi}{\partial \tau} = \Delta \pi + \pi - \pi^3 + \epsilon_z \\ \Delta \phi = \eta \dfrac{\partial \pi}{\partial z} \end{cases} \tag{3}$$

where the first one is the Landau-Khalatnikov kinetic equation and the second one is the Poisson equation with a dimensionless electric susceptibility $\eta = 1/(\varepsilon_0 \varepsilon_b |A|)$.

This review describes only the basic equations of the model, while a detailed procedure for derivation of the kinetic equations (3) and the consequent stochastic equations can be found in previous work [68]. The stochastic approach is based on introducing of two-point autocorrelation functions for the single component of polarization $K(\mathbf{s},\tau) = \langle \xi(\mathbf{r}_1,\tau)\xi(\mathbf{r}_2,\tau) \rangle$ and the electric potential, $g(\mathbf{s},\tau) = \langle \phi(\mathbf{r}_1,\tau)\phi(\mathbf{r}_2,\tau) \rangle$, with $\mathbf{s} = \mathbf{r}_1 - \mathbf{r}_2$, together with cross-correlation functions between the polarization and three electric field components, $\Psi_{xz}(\mathbf{s},\tau) = \langle \epsilon_x(\mathbf{r}_1,\tau)\xi(\mathbf{r}_2,\tau) \rangle$, $\Psi_{yz}(\mathbf{s},\tau) = \langle \epsilon_y(\mathbf{r}_1,\tau)\xi(\mathbf{r}_2,\tau) \rangle$, $\Psi_{zz}(\mathbf{s},\tau) = \langle \epsilon_z(\mathbf{r}_1,\tau)\xi(\mathbf{r}_2,\tau) \rangle$, as well as between the electric field components themselves, $R_{\alpha\beta}(\mathbf{s},\tau) = \langle \dfrac{\partial \phi(\mathbf{r}_1,\tau)}{\partial r_{1\alpha}} \dfrac{\partial \phi(\mathbf{r}_2,\tau)}{\partial r_{2\beta}} \rangle$. As will be shown in Section 3, the knowledge of correlation functions for stochastic systems is indispensable for processing experimental pictures of domain structures and evaluating macroscopic physical quantities involving products of random variables, like the energy. Analytical expressions for a complete set of polarization and electric field correlation coefficients were derived previously [69]. Regarding the practical applications, the expressions for polarization correlations are the most interesting, since only they can be so far compared with experimental data. At the same time, the study of depolarization field correlations and polarization-electric field cross-correlations are of fundamental interest and requires further experimental research.

Applying the Fourier transforms to all correlation functions $K(\mathbf{s}, \tau)$, $\Psi_{ij}(\mathbf{s}, \tau)$, and $R_{\alpha\beta}(\mathbf{s},)$ allows to reduce the number of equations and to formulate the closed system of nonlinear integrodifferential equations for the average polarization $\bar{\pi}(\tau)$ and the Fourier transform of the polarization correlation function $\widetilde{K}(\mathbf{q}, \tau)$:

$$\begin{cases} \frac{d\bar{\pi}}{d\tau} = \bar{\pi}\big(1 - \alpha_z - 3K(0,\tau)\big) - \bar{\pi}^3 + \epsilon_a \\ \frac{d\widetilde{K}(\mathbf{q},\tau)}{d\tau} = 2\left[1 - 3\bar{\pi}^2(\tau) - 3K(0,\tau) - \left(q^2 + \eta\frac{q_z^2}{q^2}\right)\right]\widetilde{K}(\mathbf{q}, \tau) \end{cases} \quad (4)$$

The solution of the system (4) requires an assumption about the random fields specified by the correlation function $K(\mathbf{s}, 0)$ at the initial time moment. This function determines, among others, the time-dependent correlation length of the domain structure. The first development of the stochastic theory was made under assumption of a Gaussian-like correlations of the initial disorder $K(\mathbf{s}, 0) \approx \exp(-s^2/(2r_c^2))$ emerged in the system immediately after its cooling into the ferroelectric phase [63, 68, 72]. The expression derived for the correlation length, $L(\tau) = \sqrt{(r_c^2 + 4\tau)/3}$ with the Gauss parameter $r_c$, described satisfactorily the available experimental data for the TGS crystals, in contrast to the longitudinal $K_{||}(\mathbf{s},\tau)$ and transverse $K_\perp(\mathbf{s},\tau)$ correlation functions [68]. The expression for the latter across the domain walls described only the general trends but, particularly, did not show the oscillations observed in the experiment, that was, in principle, expected within the framework of the isotropic model. The theoretical curve for correlations along the domains within the polar plane described the domain patterns much better, but showed the Gaussian-like spatial behavior at all times, whereas it was linear in all experiments. Since the domain evolution depends significantly on the prehistory of the sample and quenching conditions, and, accordingly, on the shape of the initial correlation function, other initial correlation function shapes might show better agreement with experiment. Therefore, the assumptions of the exponential $(K(\mathbf{s},0) \approx \exp(-s/\xi))$ and the complementary error function $(K(\mathbf{s},0) \approx \mathrm{erfc}(s/\xi))$ forms of initial correlation function were studied [70] and showed a good agreement with the pioneering experimental results by Tomita *et al.* [44] for the exponential function. Nevertheless, all the obtained forms of correlation functions can be useful for characterizing domain structures and optimized fitting, since in each specific experiment the initial disorder can have different properties as will be shown in Section 3. All the obtained expressions for correlation lengths and polarization correlation coefficients are presented in Supplementary materials S1.

The knowledge of correlations is important, since it affects the way of the domain structure evolution. Analytical treatment of the system of equations (4) leads to a system of evolutionary equations for the average polarization $\bar{\pi}(\tau)$ and its variance $D(\tau) = K(0,\tau)$ [68], which explicitly depends on the correlation length $L(\tau)$, and, thus, on the form of the initial correlation function:

$$\begin{cases} \frac{d\bar{\pi}(\tau)}{d\tau} = \bar{\pi}(\tau)\big(1 - \alpha_z - 3D(\tau)\big) - \bar{\pi}^3(\tau) + \epsilon_a \\ \frac{dD(\tau)}{d\tau} = \left[2 - 6\bar{\pi}^2(\tau) - 6D(\tau) - \frac{2}{L^2(\tau)} + \frac{1}{2\tau}\left[1 - \frac{2}{\sqrt{\pi}}\frac{\sqrt{2\eta\tau}\exp(-2\eta\tau)}{\mathrm{erf}(\sqrt{2\eta\tau})}\right]\right]D(\tau) \end{cases} \quad (5)$$

System of nonlinear equations (5) characterizes the variety of phase trajectories of the evolution of the system to the thermodynamically stable state. There are six solutions of the system (5) under asymptotic consideration at large times ($d\bar{\pi}(\tau)/d\tau \to 0, dD(\tau)/d\tau \to 0$) that make up the phase pattern (Fig. 2). Singular point I is an unstable node characterized by the initial disorder which appears immediately after quenching. After the system enters the vicinity of this point, it begins to relax to one of the thermodynamically stable states: a single-domain one directed along the applied field (point II) or against it (point III), or a multidomain state (point IV) with almost equal volumes of domains of the opposite sign (Fig. 2). There is a variety of possible phase trajectories of the domain ordering. In addition to a direct transition to points II, III, and IV, the domain structure can relax nonmonotonically through the formation of intermediate asymmetric short-lived states when the system passes in the vicinity of saddle points V and VI. Separatrices intersecting at saddle points V and VI divide the phase pattern into sectors of attraction of single-domain (*1, 2*) and multidomain (*3*) states and determine the switching region between them (Fig. 2).

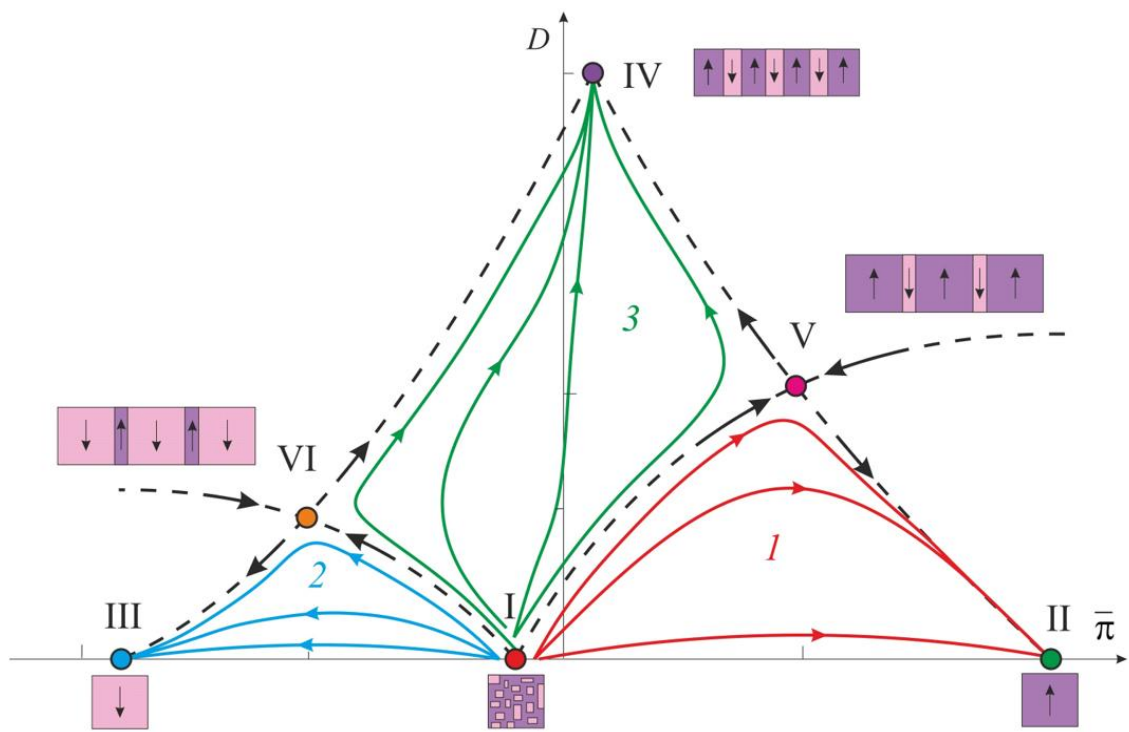

Figure 2. Schematic image of the variety of phase trajectories of domain ordering. Points I – VI are the solutions of the system of equations (5) at asymptotic case ($\partial \bar{\pi}/\partial \tau \to 0, \partial D/\partial \tau \to 0$): point I is the state of initial disorder; points II and III correspond to the stable single-domain states along and opposite to the applied field, respectively; point IV is a stable multidomain state with almost equal volume of opposite domain fractions; points V and VI are saddle points determining the short-lived phase states with pronounced asymmetry of opposite domain fractions. Dashed lines are the separatrices.

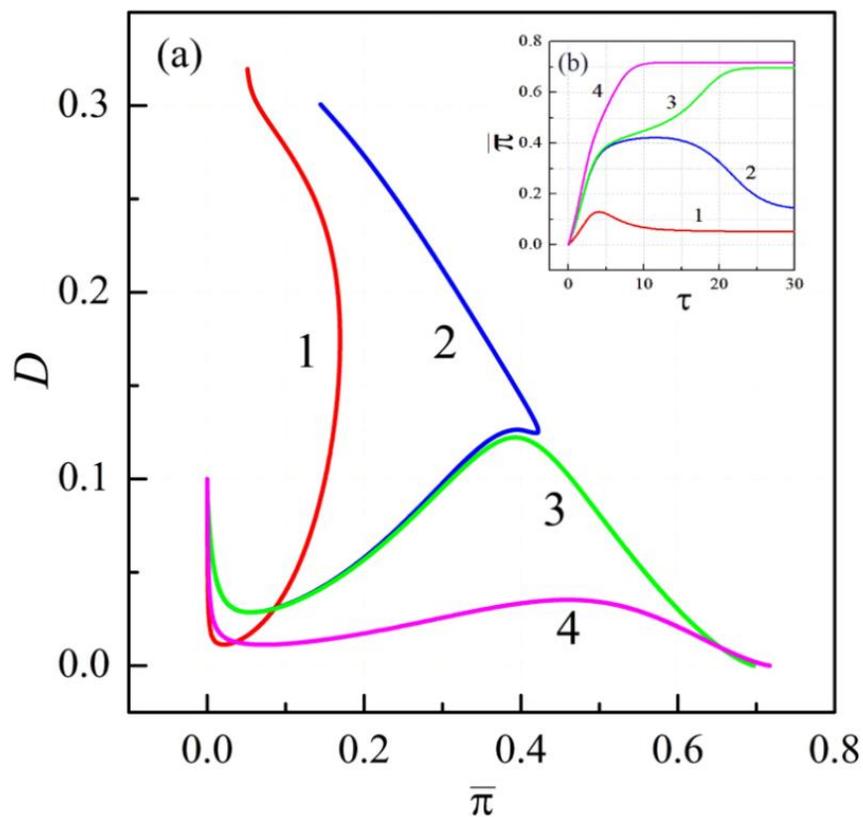

Figure 3. Numerical solution of the system of equations (5) for the initial parameters $\bar{\pi}_0 = 0$, $D(0) = 0.1$, $\alpha_z = 0.625$, $\eta = 10$, $r_c = 1$ and external electric field $\epsilon_a$: 0.03; 0.076; 0.077; 0.1 for the curves 1–4, respectively. (*a*) phase pattern in coordinates of average polarization and its variance ($\bar{\pi}$, $D$).

(*b*) – evolution curves for average polarization $\bar{\pi}(\tau)$. The solution is obtained for the initial Gaussian correlation function of polarization and the appropriate correlation length $L(\tau) = \sqrt{(r_c^2 + 4\tau)/3}$

While the asymptotical consideration of the system of equations (5) presents analytically only singular points, the numerical solution allows to describe the evolution of domain structure in detail depending on the initial conditions (initial average polarization $\bar{\pi}_0$, its variance $D(0)$, and the Gaussian parameter $r_c$), external electric field $\epsilon_a$, quenching temperature and geometry of the sample reflected by the parameter $\alpha_z$ [68, 70]. An example of a numerical solution of the system of equations (5) with an assumption of a Gaussian initial correlation function for different values of the external electric field $\epsilon_a$ imposed on the sample is presented in Fig. 3. The field affects not only the shape of the phase trajectories, but also the type of the finally formed domain structure. Thus, both curves 1 and 2 indicate the formation of a stable multidomain ordering (point IV in Fig. 2), but in the case of curve 2 the structure is more asymmetrical, because its final average polarization $\bar{\pi}$ is higher and its variance $D$ is lower than of the curve 1 (Fig. 3a). The similar situation is observed in the case of single-domain ordering: curve 4 indicates the formation of a stable domain structure with larger polarization $\bar{\pi}$ than curve 3 (Fig. 3a). The region of switching between multidomain and single-domain states is determined by the position of the saddle point V (when $\epsilon_a > 0$) and VI (when $\epsilon_a < 0$) and is presented by curves 2 and 3 in Fig. 3. The system gets in the vicinity of saddle point when the electric field imposed on the sample is about some critical value $\epsilon_a \approx \epsilon_{cr}$ assumed as a coercive field. The domain structure experiences the kinetic slowing down and stops to develop for the time which can be determined as a length of the step on the evolution curves (curves 2, 3 in Fig. 3b). However, this stage cannot be maintained infinitely, and the system will continue to relax to the thermodynamic equilibrium with the formation of a stable multidomain ($\epsilon_a \lesssim \epsilon_{cr}$) or a single-domain ($\epsilon_a \gtrsim \epsilon_{cr}$) state depending on the value of the electric field imposed on the sample (Fig. 3). It was shown that the magnitude of the coercive field changes from larger to smaller for Gaussian, exponential and complementary error function adopted for $K(\mathbf{s}, 0)$, respectively [70]. This brings us back to the importance of a detailed understanding of correlations for the use of stochastic models as a predictive and methodological tool for practical purposes in materials science and applications.

Coercive fields in ferroelectrics were derived theoretically by various methods treating the energy balance during the nucleation and growth of a domain of reversed polarization in polarized media subject to an opposite electric field. First, a classical phenomenological consideration of a large triangular reversed domain with a thin interface resulted in unrealistically large nucleation energy and the corresponding coercive field in BT [73, 74]. Later the MD and MC simulations combined with the LGD analysis allowed reasonable evaluation of coercive fields in BT and PT [40, 41]. It was shown thereby that the optimum small nuclei of the reversed domains have quadratic [40] or diamond [41] shapes with a diffuse interface. Simulations of that sort for uniaxial ferroelectrics have not been performed yet. The reviewed here LGD-based analysis in terms of stochastic characteristics of the polarization and electric fields in uniaxial ferroelectrics [68-70] does not describe a specific shape of nuclei, however, the electrostatic and domain wall energies as well as the gradual domain interface are taken into account in the LGD energy functional (1). This approach predicts coercive fields in the range from $10^{-2}$ to $10^{-1}$ of the thermodynamic coercive field [70]. This is still too large in comparison with experimental values, probably, because the domain nucleation on defects and the electrodes is not yet included in the theory that could reduce the coercive field significantly [75]. Nevertheless, the most significant result for applications here is the monotonic increase of the coercive field with both the initial amplitude and the initial spatial size of the polarization fluctuations immediately after quenching to the low-temperature ferroelectric phase [70]. Since the initial state can be controlled by, for example, initial temperature before quenching and the cooling rate [21-24,34], this presents a tool for achieving desired ferroelectric properties, represented by the coercive fields and hysteresis loops.

## 3. CORRELATIONS OF POLARIZATION

### *3.1. Correlation functions in applications*

Correlation analysis is a vital analytical tool in several practical applications: second harmonic generation (SHG) and scattering (SHS), diffuse, neutron and dynamic light scattering, piezoelectric (PFM), scanning (SFM) and atomic force (AFM) microscopy [44 – 48, 51, 54 – 56, 76 – 90]. Correlation functions arise explicitly when structures that are inhomogeneous in space are studied by SHG and SHS [91]. It provides information about the symmetry, regularity and polarization orientation distribution in ferroelectric samples. The modulation function was introduced for measuring the mean domain thickness and change of the second harmonic intensity coinciding with the polarization correlation function assuming +1 in positive domains, -1 in negative domains and 0 in domain walls or outside the crystal [76]. Since the piezoresponse signal is proportional to the local polarization value, a polarization-polarization correlation function is used over years for describing domain pictures obtained via PFM and provides detailed information about characteristic parameters such as the mean domain size, correlation length and period of the structure [77 – 87].

Evaluation of the correlation function with the fast Fourier transform has proven to be very useful for assessing the characteristics of a domain structure in thin films and epitaxial strained heterostructures [87, 89]. This method is more sensitive to the period distortions of the surface of the sample than the direct fast Fourier transform of the topography image [88, 92]. Another useful property of correlation analysis is the ability to process and compare domain patterns obtained by different methods. The radially averaged autocorrelation function showed the characteristic correlation length of 800 nm for SHG contrast variations and only 100 nm for the PFM images corresponding directly to the domain structure of lead titanate (PbTiO$_3$) thin films [88]. It allows to provide further blurring of PFM images to mimic lower resolution and averaging effect of SHG plots and clearly reveal the long-range variations in density of domains.

The characteristic size ("maze period") used as the statistical characteristic of the maze domain structure was defined as the position of the first maximum of the autocorrelation function, and was found to be 120 ± 40 nm in SBN single-crystals [77]. In lithium niobate (LiNbO$_3$) crystals placed between two electrodes the average domain size (0.8-2 µm), structure period (4 µm) and depth (8 µm) of the quasi-regular chain of rounded domains was obtained by the autocorrelation analysis [78]. In the PIMNT27 (Pb(In$_{1/2}$Nb$_{1/2}$)O$_3$-0.49Pb(Mg$_{1/3}$Nb$_{2/3}$)O$_3$-0.27PbTiO$_3$) and PMNT27 (Pb(Mg$_{1/3}$Nb$_{2/3}$)O$_3$-0.27PbTiO$_3$) single-crystals the values of mean short-range correlation length of poled nanoscaled domain structures were determined as ~80 nm and ~151nm, respectively [79].

Correlation functions are especially useful for analyzing the domain structure in ferroelectric ceramics that exhibits the multi-type domain patterns comprising irregular island domains and regular lamellar domains at the nanoscale. In this way the mean domain size was estimated in PIN-PMN-PT (Pb(In$_{1/2}$Nb$_{1/2}$)O$_3$-Pb(Mg$_{1/3}$Nb$_{2/3}$)O$_3$-PbTiO$_3$) ceramics depending on PIN proportion: 205 nm, 154 nm and 65 nm for 6PIN, 24PIN and 58PIN samples, respectively [80]. The regular in two different directions autocorrelation function of PFM images for relaxor PLZT (Pb$_{1-x}$La$_x$Zr$_{1-y}$Ti$_y$O$_3$) ceramics revealed the local rhombohedral symmetry [84]. The average correlation radius was estimated as ~50 nm and the correlation length as 80-90 nm. This knowledge is extremely important to control the properties of relaxors, because the higher the correlation length, the lower the degree of disorder in the material which is possible to map via PFM [84]. The two-point polarization-polarization correlation function $K(\bm{r},t) = \langle S(\bm{r},t)S(0,t) \rangle$ representing the polar disorder in uniaxial ferroelectrics, with $S$ taking the values +1 in positive and -1 in negative domains [44 – 48, 54 – 56] can adopt different mathematical forms. The Gaussian-like function $\langle K(s) \rangle \sim \exp(-(s/\langle r_c \rangle)^{2b})$ was used by experimentalists [79, 84] where $s$ is the distance from the central peak; $\langle r_c \rangle$ is the average short-range correlation length related to the mean size of polar nanodomains; $b$ is the parameter of polarization interface roughness. The exponential function $\langle K(s) \rangle \sim \exp[-(s/\langle \xi \rangle)^{2h}]$, where $0 < h < 1$ was used to evaluate correlation characteristics in the PGO crystal [34]. A more complicated form of correlation function including both short-range and long-range contributions showed a better fitting with PFM images in Refs. [81 – 83]

$$K(s) = \sigma^2 \exp\left[-\left(\frac{s}{r_c}\right)^{2b}\right] + (1-\sigma^2)\exp\left[-\frac{s}{\xi}\right]\cos\left(\frac{\pi r}{a}\right), \qquad (6)$$

where $\xi$ is a long-range correlation length and $a$ is a period of nanodomain structure, $\sigma$ is a dimensionless fitting parameter. Here, the first term of the expression (6) describes the short-distance correlations of the polarization piezoresponse signal inside the individual nanodomain and the second term determines the long-range correlations reflecting the regularity of the nanodomain pattern. Thus, the shape of the correlation curve is not limited to the expressions proposed previously [70], and more complex mathematical expressions may be required to adequately characterize domain patterns. We note, however, that the time-dependent correlation function resulting from the theory [68 – 70] does not retain its initial shape in the course of the evolution.

### *3.2. Polarization Correlations in TGS single-crystals*

In the pioneering papers [42 – 45] the domain kinetics in TGS single-crystals was observed by the nematic liquid crystal method employing the conventional optical microscopy for domain detection. Although the spatial resolution of a polarized light microscope is limited by diffraction, a qualitative analysis of obtained images showed the smooth curves of both longitudinal and transvers correlations of polarization in the polar plane. Further experimental studies were performed by SFM [46 – 48] and AFM [54 – 56] using the voltage modulation techniques that allowed to avoid the liquid crystal coating on the ferroelectric sample.

With the progressive development of experimental methods for observing the domain structure in TGS, theoretical processing of domain patterns was still carried out using the simplest approach described in the pioneering paper [44]. Domain images were converted to the binary matrix consisting of the only +1 and -1 values, corresponding to the white (positive domain) and black (negative domain) regions thus neglecting the transition areas between such regions, i.e. domain walls. In the [100] crystallographic direction, i.e. approximately along domain walls (up to 16º angle due to the anisotropy of TGS within the polar plane), the correlation coefficient $C(s)$ represents a correlation of the first column of the matrix with all others. Similarly, in the [001] direction, i.e. perpendicular to the domain walls, $C(s)$ gives correlations between the first line of the matrix with the others [67]. The correlation length $L(t)$ was determined as the distance *s* where the absolute value of correlation coefficient is one half $C[s = L(t), t] = 1/2$ [47]. The disadvantage of this method is the need to process small fragments of the visual area (0.02×0.02 mm$^2$ from the available 0.3×0.3 mm$^2$), otherwise the correlation curves manifest large oscillations, especially in [001] direction.

It was assumed by the authors of experiments [54 – 56] that the smooth shape of correlation curves in the previous works [44 – 48] resulted from the polynomial approximation of initial data. In turn, we assume that significant irregular oscillations of experimental correlation curves in Refs. [54 – 56] are due precisely to the specifics of the above described matrix treatment. As was shown before [70] such domain image processing does not account for the finite domain wall thickness that can affect the form of the correlation curves, especially at small distances $s$.

Analytical expressions of the correlation coefficients derived by the authors of this study [68, 70] were used to fit the pioneering experiment by Tomita *et al*. [44]. It was shown that, in the isotropic approximation, only the formula for the correlation coefficient along the domains in the polar plane is appropriate for the description of correlation, while the expression for correlations perpendicular to the domain walls does not fit the oscillation behavior [68]. This limitation arises from the form of the LGD functional, neglecting the crystal anisotropy within the polar plane which is the reason of the observed lamellar domain shape. Subsequently, by exploration of various forms of the initial correlation function (Gaussian, $K(s,0) \approx \exp(-s^2/r_c^2)$, exponential, $K(s,0) \approx \exp(-s/\xi)$, and complementary error function $K(s,0) \approx \text{erfc}(s/\xi)$), analytical expressions were derived for correlations along domains [70] and summarized in Supplemental materials S1. Their application to the experiment by Tomita *et al*. [44] showcased a preferred fitting using the exponential $K(s,0)$ shape; nevertheless, a validation against additional experiments remains imperative to substantiate or rebut this assertion.

In the experimental papers by Likodimos *et al.* [46 – 48] the only correlation function perpendicular to the domain walls has been shown, which cannot be fitted good within the outcomes of our model. Anyway, they all can be used to evaluate the expressions for the correlation length obtained [70]. It can be seen that both exponential and complementary error functions show good fitting of the experiment [46], but not the Gaussian one (Fig. 4).

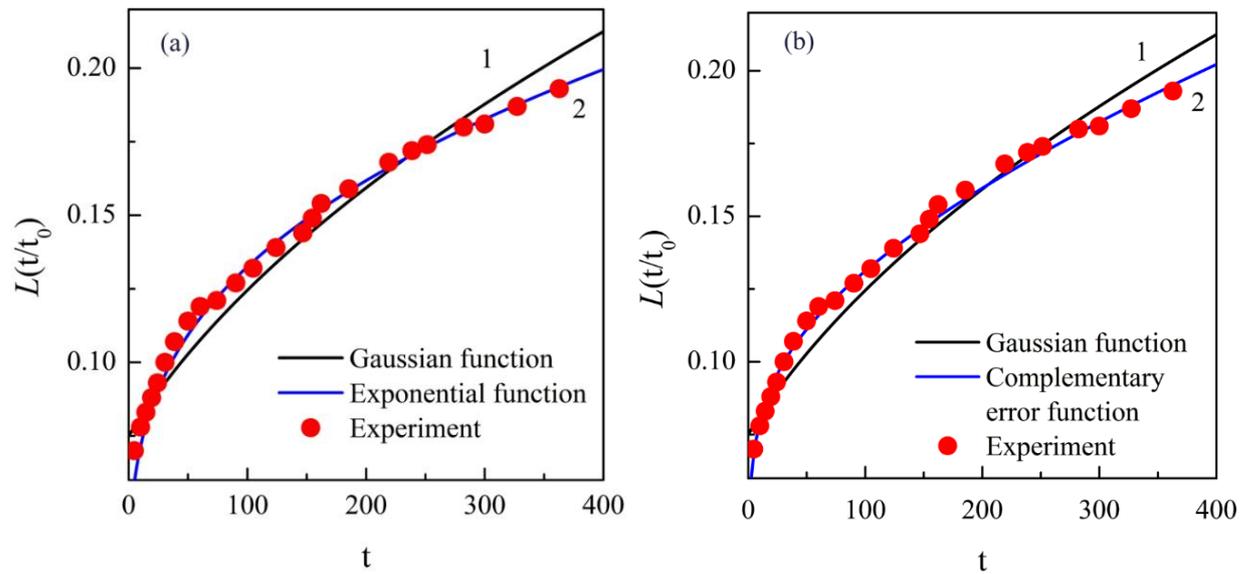

Figure 4. Fitting the experimental data (dots) for correlation length $L(\tau)$ from Likodimos *et al.* [46] (a) for the Gaussian (curve 1) and exponential (curve 2) initial correlation functions; (b) for the Gaussian (curve 1) and the complementary error (curve 2) functions

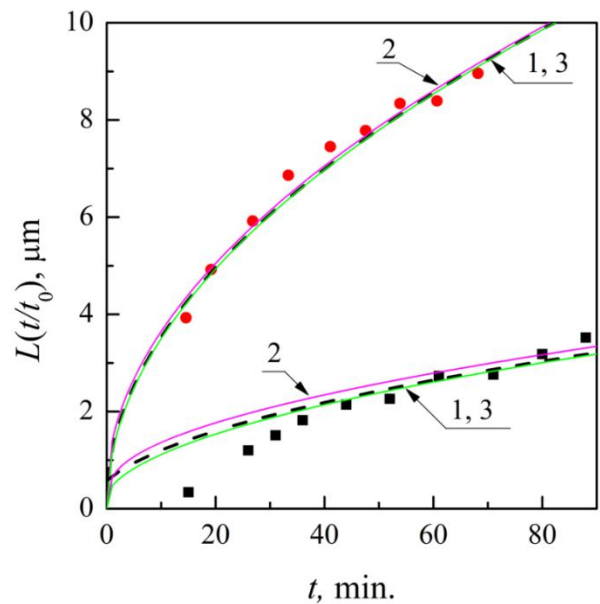

Figure 5. Fitting the experimental data for the correlation length $L(\tau)$ from Golitsina *et al.* [55] along the lamellar domains (circles) and across the lamellar domains (squares) by different initial

correlation functions expressions: Gaussian (curves 1), exponential (curves 2) and complementary error function (curves 3) [70]. Gaussian and erfc functions are almost merged

In experiments made by Golitsina *et al.*, correlation functions and corresponding correlation lengths were assessed in the polar plane [55]. However, endeavors to pinpoint optimal fitting parameters proved less fruitful compared to prior instances [70]. Owing to considerable scattering among experimental data points, all three analytical expressions for $L(t)$ were deemed suitable only with a notable margin of error, particularly evident in the direction across lamellar domains (Fig. 5). Such discrepancies were anticipated, given the methodology adopted by experimenters to derive the correlation length from the correlation function, which exhibits a smoother profile in the Likodimos experiment relative to the Golitsyna experiment.

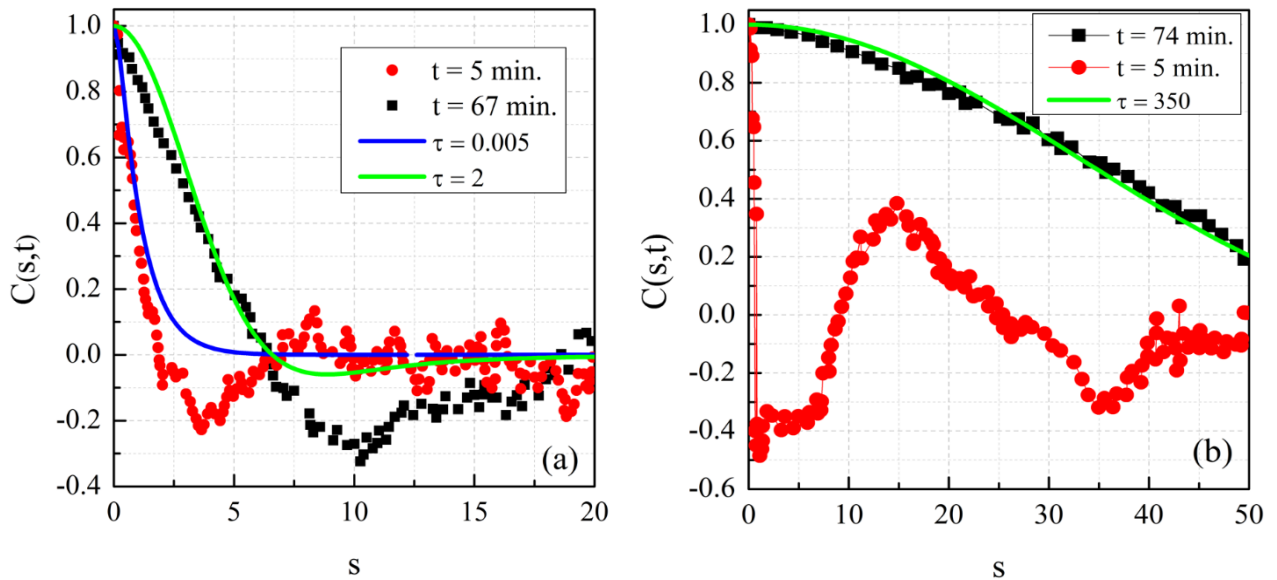

Figure 6. Fitting of experimental data on correlations along the lamellar domains by Golitsina *et al.* [55] at quenching temperature $\Delta T$ of 0.1 K (a) and 4 K (b) with exponential correlation coefficient [60] for $\tau = t/t_0$ with $t_0 = 35000$

In the experiments conducted by Golitsina *et al*. [55], correlation coefficients were obtained both along the longitudinal and transverse directions with respect to domains in the polar plane for the pure, doped, and irradiated TGS crystals at various quenching temperatures. Theoretical curve comparisons facilitated by employing the correlation coefficients derived from the assumed exponential form of the initial correlation function [70] along the lamellar domains, demonstrated satisfactory concordance with experimental findings. Notably, when the sample underwent quenching close to the Curie temperature ($\Delta T = 0.1$ K), distinct oscillations were observed, defying quantitative description within the confines of the isotropic model, evident over the short and long timescales (Fig. 6a). Deeper quenching conditions ($\Delta T = 4$ K) similarly revealed pronounced spatial linearity in correlation curves at shorter times, yet exhibited significant alignment with theoretical predictions only at extended times (Fig. 6b).

The results obtained show that the improvement of both experimental domain images processing and the stochastic theory taking into account the influence of the crystal anisotropy are on the agenda. This approach enables a more comprehensive description of not only kinetic experiments but also puzzling phenomena, such as the absence of charge on head-to-head (H-H) and tail-to-tail (T-T) domain walls observed in TGS and PGO [93]. Previously we derived the charge density correlation functions from polarization correlation functions as $M(\boldsymbol{s}, \tau) = -\partial^2 K(\boldsymbol{s}, \tau)/\partial s_z^2$ [70] and found that the charge density correlations and fluctuations are significantly suppressed. This suggests that the actual charge on domain walls is an order of magnitude lower than expected for H-H and T-T configurations in uniaxial ferroelectrics. Since our model does not contain free charge carriers, this means that the bound charges must be effectively reduced by one order of the magnitude by some structural organization of domains, which can be, in particular, the saddle-point domains structures observed by McCluskey et al. [93].

*3.3. Evolution of polarization correlations in time and space*

The advanced stochastic theory not only enables the description of existing experimental data but also facilitates the prediction of the temporal and spatial dependence of correlations. The analytical expressions for the polarization correlation coefficient, denoted as $C(\mathbf{s},\tau)$, were derived in both precise and approximate forms for the initial Gaussian correlation function [68, 69], incorporating the correlation length $L(\tau)$ and contingent upon the correlation radius $r_c$ and the susceptibility parameter $\eta$. Prior investigations in [69] were performed for parameters $s = 3$, $r_c = 1$ and $\eta = 1$, without accounting for the dynamic evolution of polarization correlations across time and space. Hereinafter, the correlation coefficients depending on space and time are calculated by formulas written in spherical coordinates and systematized in Supplemental materials S1.

The polarization correlations are isotropic in the polar plane and depend only on the distances between random points denoted as $s$ and the polar angle the vector $\mathbf{s}$ makes with the positive z-direction. The temporal progression of polarization correlations is depicted in Fig. **7** as a function of the polar angle and time for various fixed distances between random points. When the points are situated closely, i.e. at a distance $s$ less or about $L(\tau)$, the correlations between them are noticeable (~ 0.5 in dimensionless units) and isotropic, even at very small times (Fig. 7a). This phenomenon can be rationalized by positioning of the points within a single domain characterized by a consistent direction of the polarization vector: the closer the points are (smaller $s$), the higher the probability of their coexistence within the same domain. Consequently, polarization correlations manifest themselves isotropically and swiftly evolve with time towards a constant value (Fig. 7a). Conversely, as the distance $s$ increases, the correlations weaken and evolve at a slower pace (Fig. 7b). Initially, these correlations display insignificant amplitude (~ 0.05 in dimensionless units) and exhibit a clear orientation along the polar axis of the crystal ($\theta = 0$). They increase much slower than in the case of small $s$, and one can observe the transition from anisotropic to isotropic shape (Fig. 7b). It can be assumed that this behavior is related to the growth of the domain structure, and when both points are within the same domain, the polarization correlations become isotropic.

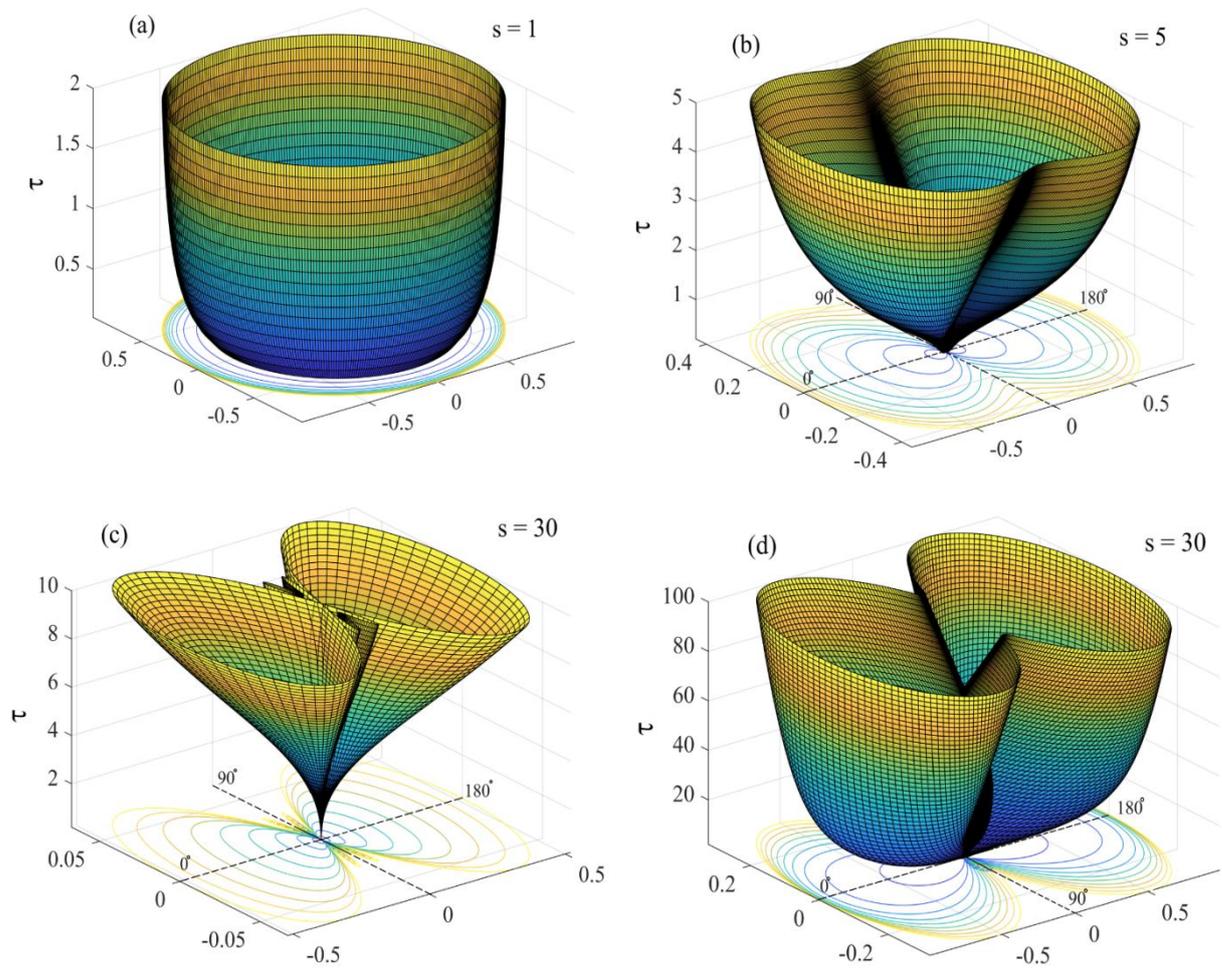

Figure 7. Time evolution of polarization correlations $C(\mathbf{s}, \tau)$ at different distance between random points *s* for the parameters $r_c = 1$ and $\eta = 1$. Plots (c) and (d) are built for the same s = 30 but in different time scale

As the distance between points increases, Fig. 7c illustrates a notable trend: the correlations diminish in amplitude and exhibit increasing anisotropy. This evolution occurs at a sluggish pace, with the anisotropic characteristics becoming apparent not solely along the polar axis, but also in more exotic directions which reveals the domain superstructures in the polar plane. Such behavior finds explanation in the proposition of numerous small domains existing between two widely spaced random points. Over time, even these correlations grow towards isotropy; however, this transition requires a considerably longer duration compared to scenarios involving smaller distances $s < 10$ (Fig. 7d).

Figure 8 illustrates the progression of correlations, as the distance between points increases, showcasing both the attenuation of correlations (a) and the accentuation of anisotropic tendencies (b), notably along the polar axis and with subtle manifestations in other directions. Initially, at small times, correlations persist isotropic despite increasing point-to-point distances (a) but rapidly diminish in amplitude. Subsequently, polarization correlations exhibit a slower attenuation with the increasing point-to-point distance (b). Notably, a pronounced anisotropy along the polar axis emerges, accompanied by faint "noise" in its proximity (b).

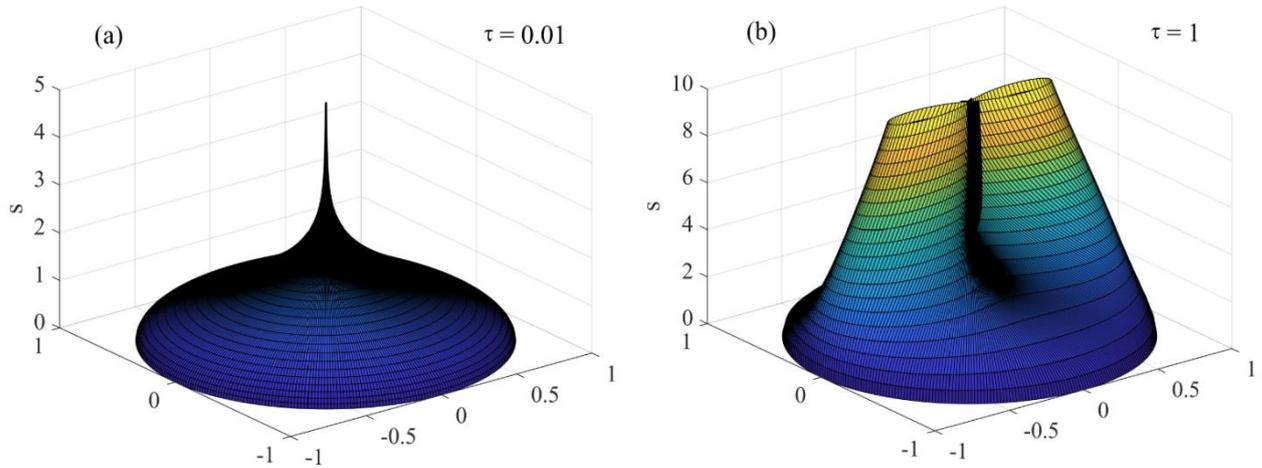

Figure 8. Evolution of polarization correlations $C(\mathbf{s}, \tau)$ in space for the parameters $r_c = 1$ and $\eta = 1$ at different moments of time: $\tau = 0.01$ (a); $\tau = 1$ (b)

Special consideration should be devoted to investigating the dependence of polarization correlations on the dimensionless susceptibility $\eta$ and the Gauss parameter $r_c$, giving the scale of the initial isotropic correlations. The augmentation of $r_c$ fosters the isotropic characteristics of polarization correlations (Fig. 9a). With sufficiently elevated values of the parameter $r_c$, these correlations exhibit substantial amplitude and remain constant over time, akin to scenarios involving small distances between points (Fig. 7a).

Parameter $\eta = 1/(\varepsilon_0\varepsilon_b\alpha_0|T - T_C|)$ defines not only the specific properties of material, but also their temperature dependence. According to the known values for TGS crystal, $\varepsilon_b \approx 7.5$, $\alpha_0 = 1.5 \cdot 10^{10}$ J·m/C$^2$, and $\varepsilon_0 = 8.85 \cdot 10^{-12}$ F/m, the temperature dependence can be assumed approximately equal to $\eta \approx 1\text{K}/|T - T_C|$. The smaller the value $\eta$, the far the system from the phase transition temperature. All previous pictures were obtained for the values $\eta = 1$, which is away from $T_C$, and $\eta = 10$, which is close to $T_C$ but out of the region of critical thermal fluctuations, and showed good fitting for $\eta = 10$ with known experimental results [70]. But when the system is far from phase transition temperature, say, $|T - T_C| = 10$ K, i.e. $\eta = 0.1$, the correlations of polarization become completely isotropic and just increase monotonously with time (Fig. 9b).

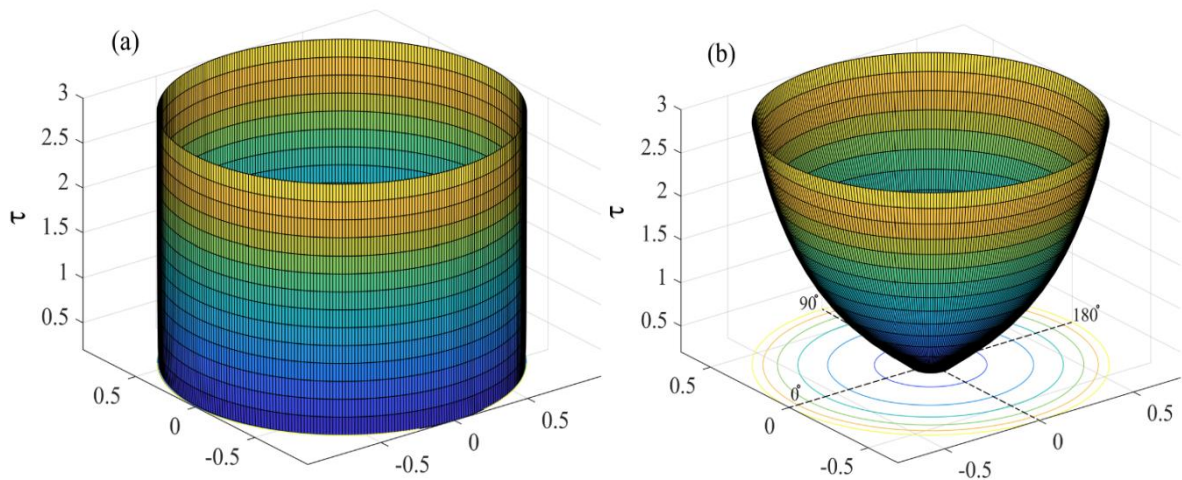

Figure 9. Time evolution of polarization correlations $C(\mathbf{s}, \tau)$ at $s = 3$ for parameters: (*a*) $r_c = 10$, $\eta = 10$; (*b*) $r_c = 1$, $\eta = 0.1$

To summarize, the temperature dependence of polarization correlations through the parameter η determines only their amplitude, but not their shape, since over the entire temperature range $|T - T_C|$ they are isotropic (Fig. 9b). Anisotropic behavior appears only with increasing distance between points *s* (Figs. 7, 8).

## 4. ELECTRIC FIELD CORRELATIONS

The advanced stochastic approach has enabled the derivation of analytical expressions not only for polarization correlations but also for more intricate correlations, such as electric field correlations and electric field-polarization cross-correlations [69], which remain so far beyond experimental detection capabilities, though, in principle, this were possible by means of the electron-holographic tomography [94]. Prior investigations have demonstrated anisotropic characteristics of such correlations in polycrystalline perovskite ferroelectrics [95], as well as in TGS single crystals within the framework isotropic in the polar plane [69], albeit without delving into their temporal and spatial evolution.

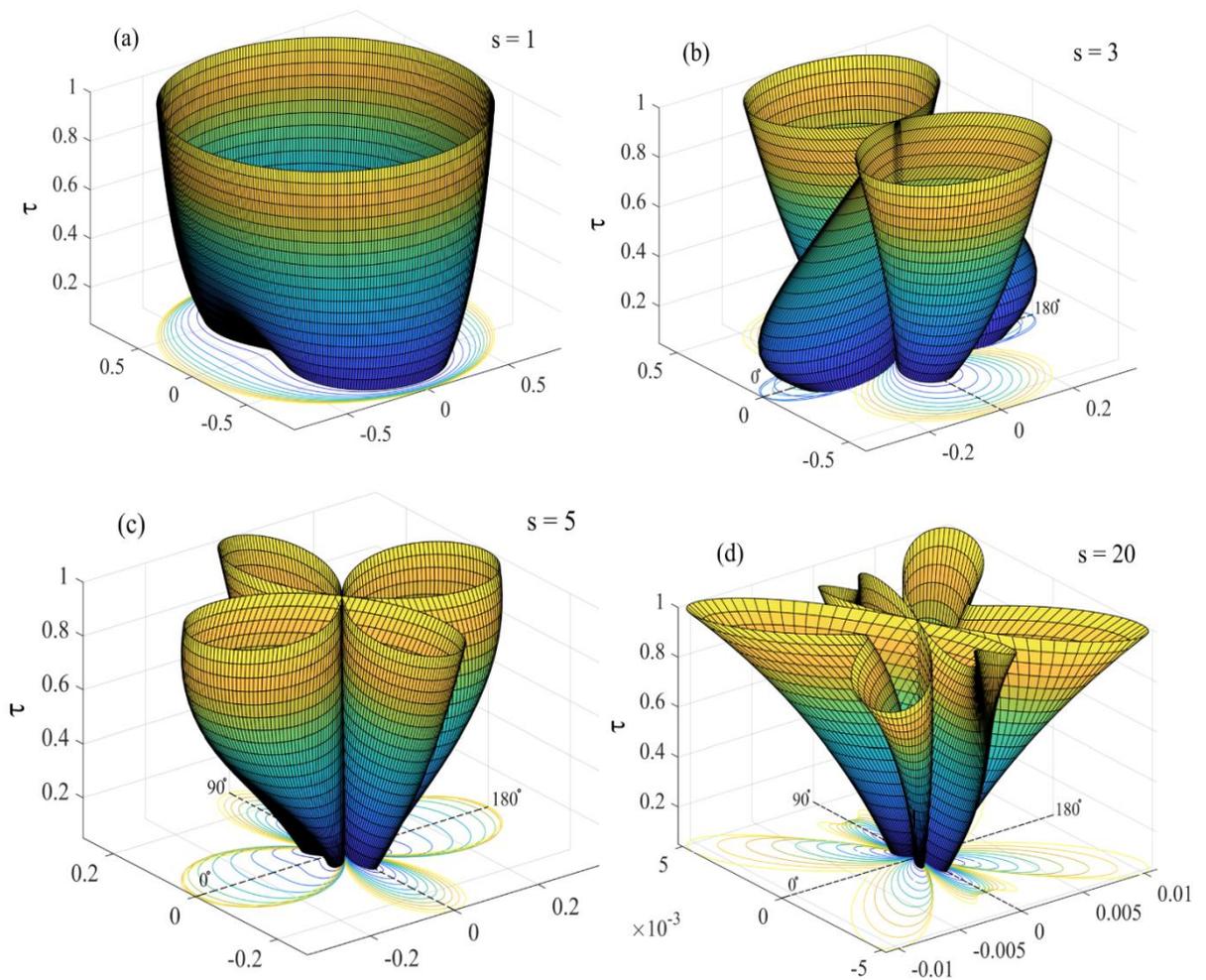

Figure 10. Time evolution of electric field correlations $r_{zz}(\mathbf{s}, \tau)$ depending on distance $s$ for parameters $r_c = 1$, $\eta = 1$. Plots (a – d) correspond to the values $s$: 1; 3; 5; 20. The detailed time evolution of correlations in time for $s = 3$ and $s = 20$ is presented in movie in supplemental materials S2(a,b)

### 4.1. Cylindrically symmetrical correlations

As well as for the polarization correlations, the autocorrelation coefficient of the electric field components along the polar axis $r_{zz} = R_{zz}(s,)/R_{zz}(0,\tau)$ is cylindrically symmetrical and shows similar behavior, but the anisotropic features are more pronounced (Fig. 10). When the distance is small, $s = 1$, the correlations have a weak anisotropy in the direction perpendicular to the polar axis, but it disappears very fast with time (Fig. 10a). At $s = 3$, correlations initially manifest themselves along the polar axis, gradually transforming to correlations perpendicular to it with a slower progression towards isotropy afterwards (Fig. 10b). When $s = 5$, correlations emerge simultaneously along and across the polar axis, exhibiting similar amplitudes and only marginal differences in shape, particularly notable at shorter durations (Fig. 10c). With an increase in the distance between two random points, electric field correlations diminish in amplitude while adopting more intricate patterns which become especially noticeable when $s = 20$. Previous findings indicated the absence of correlations along the polar axis in this scenario [69]; however, a detailed examination of their evolution reveals their gradual emergence with time, albeit less pronounced compared to other directions (Fig. 10d). Consequently, even in this instance, correlations will eventually evolve towards an isotropic state, albeit over an extended timeframe. Movie in Supplemental materials S2 presents the detailed evolution of these correlations in time (for $s = 3$ (S2a) and $s = 20$ (S2b)) and space (for $\tau = 0.01$ (S2c)).

### *4.2. Correlations without cylindrical symmetry*

Analytical expressions for all correlation coefficients described below are derived in spherical coordinates on a sphere of radius $s$ and presented in Supplemental materials S1. Correlations of the electric field components along the non-polar axes $x$ and $y$ lack cylindrical symmetry, necessitating their depiction through static images for specified values of $s$ and $\tau$. These images were visualized in Cartesian coordinate system within the Matlab package (R2023b). The evolution of such correlations in time and space can be presented only via video presentations (refer to movies in Supplemental materials S3 and S4).

Similar to earlier instances, an atypical anisotropic configuration of electric field correlations along the *x*-axis $r_{xx}=R_{xx}(\mathbf{s},\tau)/R_{xx}(0,\tau)$ is discernible for $s > 3$ and in the early stages of evolution, showcasing an enhanced sensitivity to the parameter selection (Fig. 11). The perplexing shape of these predicted correlations presents the way for further investigations and underscores their significance in ongoing research endeavors. Movie in Supplemental materials S3 presents the detailed evolution of $r_{xx}$ correlations in time (S3a) and the rotation movies present them in all angles for $s = 3$ at different time moments $\tau = 0.01$; 0.1 and 1 (S3b – S3d).

In contrast to $r_{xx}$, correlations, which manifest themselves in a certain way both in exotic directions and along and across the polar axis, the $r_{xy}=R_{xy}(\mathbf{s},\tau)/\sqrt{R_{xx}(0,\tau)R_{yy}(0,\tau)}$ correlations avoid the polar axis for any values of *s* and $\tau$ (Fig. 12) [69]. However, the main trend regarding the weakening of the amplitude of correlations and an increase in their anisotropy with increasing *s* remains (Fig. 12b). The animated plots of correlations $r_{xy}$ for the $s = 3$ at times $\tau = 0.01$; 0.1; 1 (S4a – S4c) and $s = 5$ at $\tau = 0.1$ (S4d) are presented in Supplemental materials S4.

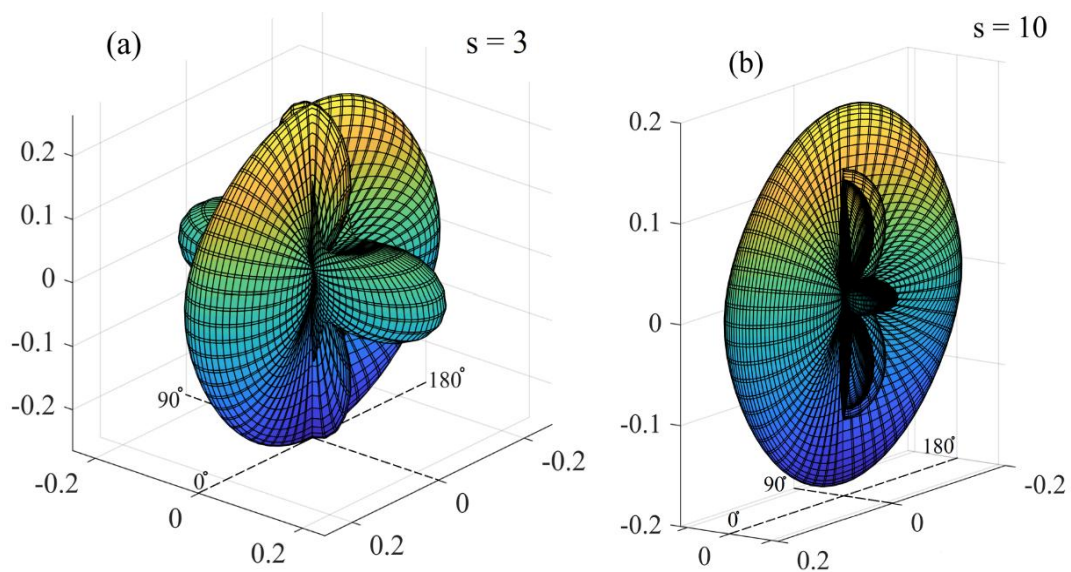

Figure 11. Electric field correlations $r_{xx}(\mathbf{s},\tau)$ for parameters: (*a*) $\tau = 0.01$ and $s = 3$; (*b*) $\tau = 1$ and $s = 10$. Animated plots are presented by movies in Supplemental materials S3

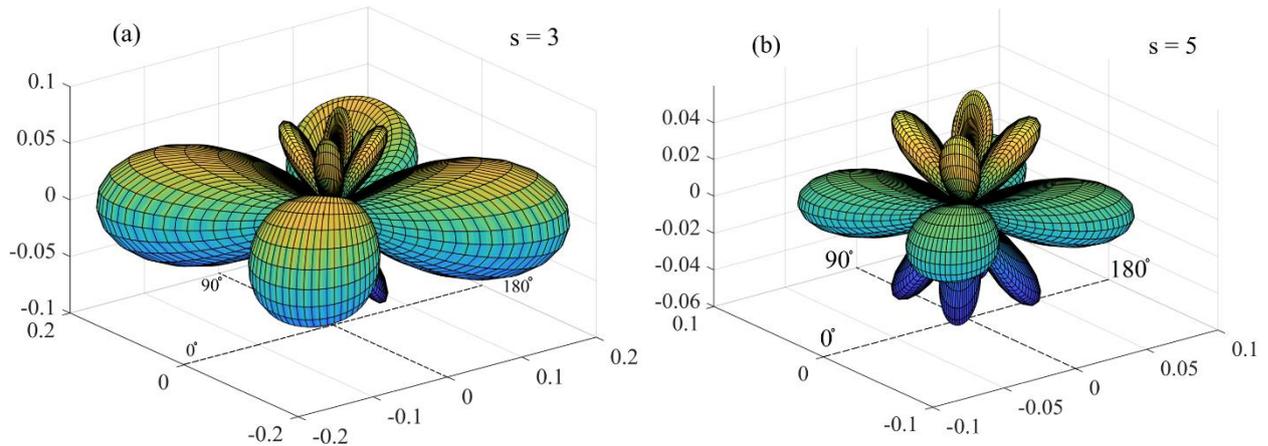

Figure 12. Electric field correlations $r_{xy}(\mathbf{s},\tau)$ for parameters: (*a*) $\tau = 0.01$ and s = 3; (*b*) $\tau = 1$ and *s* = 5. Animated plots are presented by movies in Supplemental materials S4

## 5. CROSS-CORRELATIONS BETWEEN POLARIZATION AND ELECTRIC FIELD COMPONENTS

Formation of finite-size domains in a uniaxial ferroelectric creates local bound charges which affect other domains via emerging electric depolarization fields. That is why the cross-correlations between polarization and electric field are expected to play a significant role in the domain structure formation and its kinetics. And in fact, as it was shown previously [68], taking into account the correlations between the field and polarization promotes the formation of a multidomain structure and enhances by a factor about three the coercive field which triggers the development towards the single-domain state.

Analytical expressions for cross-correlations between polarization and electric field $\psi_{zz}(\mathbf{s},\tau) = \Psi_{zz}(\mathbf{s},\tau)/\sqrt{D(\tau)R_{zz}(0,\tau)}$ were previously obtained [69]. They exhibit cylindrical symmetry as in the case of the polarization correlation coefficient $C(\mathbf{s},\tau)$ and show a similar temporal evolution (Fig. 7), but much less pronounced (cf. Fig. 7c and 13c). Strong isotropic correlations at the initial stage weaken very quickly with distance, but acquire unusual anisotropic shapes with a pronounced directionality along the polar axis (Fig. 14). The detailed time evolution of these correlations for $s = 3$ and the development in space at $\tau = 1$ is shown by movies S5 in Supplemental materials S5 (S5a and S5b respectively).

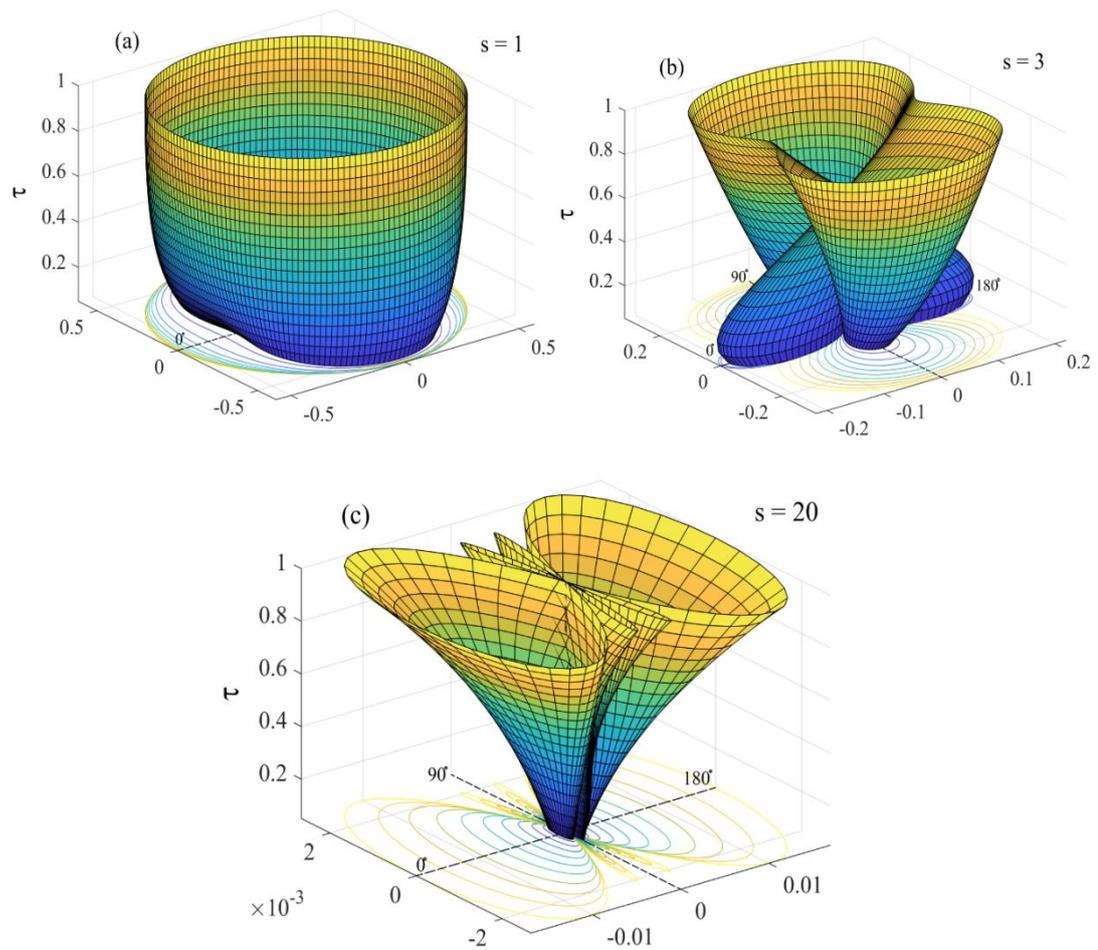

Figure 13. Time evolution of $\psi_{zz}(\mathbf{s}, \tau)$ cross-correlations as a function of the polar angle at different values $s$: 1 (a), 3 (b), 20 (c). The detailed development of correlations is shown by movie S5a in Supplemental materials S5

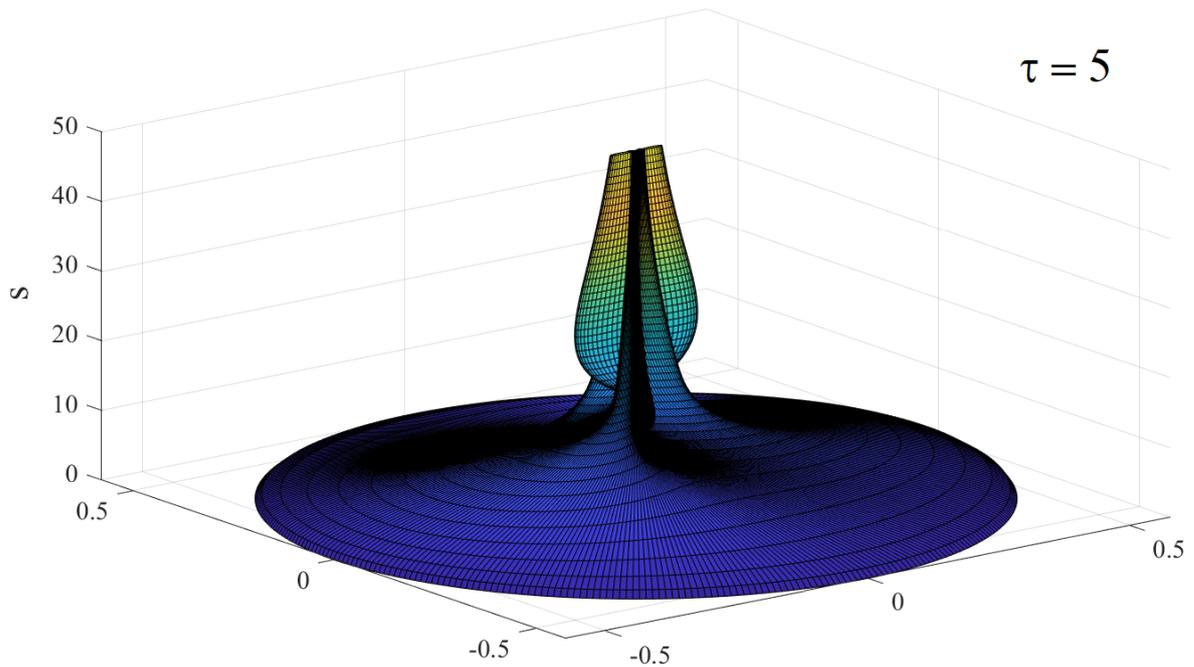

Figure 14. Evolution of $\psi_{zz}(\mathbf{s}, \tau)$ cross-correlations as a function of the polar angle depending on distance *s* at $\tau = 5$. It is presented in detail by movie S5b for $\tau = 1$ in Supplemental materials S5

At present, the developed stochastic theory offers a descriptive examination of polarization and electric field correlations, relying on simplistic assumptions regarding their exotic behavior. However, to attain a more profound comprehension of the domain structure ordering process, it becomes imperative to juxtapose the model against both experimental and simulation data.

# 6. CONCUSION: CURRENT CHALLENGES IN KINETIC STUDIES OF FERROELECTRIC DOMAIN STRUCTURES

Insights into the processes of domain structures formation unveil extensive possibilities for innovating ferroelectric materials and enhancing their functional properties [18]. However, currently kinetic studies experience numerous difficulties across experiments, simulations, and theoretical investigations.

*Experiments*. At present, the most detailed observations of domain ordering have been carried out on TGS crystals, where the structure evolution unfolds over several hours, allowing for detailed examination by means of SFM and AFM methods [46 – 48, 51 – 56]. This feasibility stems from TGS intrinsic characteristics as a "clear" dielectric boasting high insulating properties, thus minimizing distortions arising from space charge accumulation, electronic and ionic conduction, and other screening effects. For comparison, the lead germanate $Pb_5Ge_3O_{11}$ (PGO) crystal is uniaxial and has a simple 180º domain structure similar to TGS. However, its dual properties of ferroelectricity and semiconductivity, although advantageous in electro- and acousto-optics, significantly affect the nucleation, growth and dynamics of domains, limiting kinetic observations to only seconds. Presently, only static piezoresponse force microscopy (PFM) images for PGO have been obtained [34, 96], unveiling intriguing domain configurations such as uncharged H-H and T-T domains with saddle points allegedly housing accumulated mobile charges. However, their temporal evolution remains unexplored.

In contrast to PGO the archetypical perovskite ferroelectric barium titanate, $BaTiO_3$ (BTO), shares similar insulating properties with TGS, but it undergoes several first-order phase transitions. Upon cooling below the Curie temperature, BTO exhibits a sharp polarization jump, rapidly transitioning to a polar state even with a relatively gradual temperature variation. Observation of polarization kinetics in BTO presents substantial challenges due to its complex domain structure variation, stemming from three phase transitions and the presence of various domain wall types. Optical microscopy cannot distinguish the 180º domains since the change of optical indices is only related to the strain and not direction. Even though the etching of the BTO sample with hydrochloric acid allowed to observe both 90º and 180º domains this method is destructive and not useful for dynamic studies [97]. Observation of 180º domains without 90º ones is also possible if the crystal is exposed to a weak electric field normal to the *c*-axis long enough otherwise their relaxation is a fraction of a second [98 – 101]. With 90º domain walls, the nucleation requires a higher field than the movement and therefore the walls move rapidly to an equilibrium position once they have been formed. The *in-situ* observations of domain evolution in BTO as a function of temperature were carried out using a polarizing optical microscope in conjunction with a charged-coupled device camera system [97]. The evolution of domain structure in BTO was studied during heating and cooling in a wide temperature range at small speed 2 K/min, but even this was not enough to observe the detailed kinetics of ordering process [101].

*Simulations.* When technological constraints impede experimental observation of domain structures, simulation methods emerge as invaluable tools. Techniques such as atomistic modeling, density functional theory and phase-field approach enable the investigation of domain formation under diverse conditions, conserving laboratory resources and simulating experiments that may be unfeasible otherwise [30 – 38, 102 – 104]. However, simulation studies encounter the converse complexity: while they adeptly describe the behavior of multiaxial ferroelectrics like perovskites, they meet difficulties when applied to uniaxial crystals. The primary challenge in describing, for example, TGS by phase-field modeling lies not in its domain structure complexity, but rather in accurately accounting the anisotropic features associated with its monoclinic crystal symmetry. Specifically, the phase-field model for TGS would need to include terms that account for the anisotropic elastic energy, domain wall energy, and electrostatic properties associated with the monoclinic crystal structure. The model parameters would need to be calibrated based on experimental observations to ensure that the simulated domain structures and their dynamics match those observed in real TGS crystals.

Although the timescales of phase-field simulations are significantly shorter (from picoseconds to nanoseconds) than the domain structure evolution in real crystals, which span from fractions of a second in perovskites to hours in TGS, they still provide meaningful insights. While direct comparison is challenging, these simulations allow us to predict and visualize final domain structures based on the initial state of the system and external influences such as temperature, electric field, or pressure. The obtained results can be aligned with the asymptotic solutions of the evolution equations (5), i.e. the stable points II–IV in Fig. 2. Moreover, phase-field simulations capture the stepwise evolution of domain structures, enabling numerical analysis via the autocorrelation function. The obtained correlation curves can be compared with analytical ones, bridging the gap between simulation, theory and experiment. This makes phase-field modeling a valuable tool for assessing the reliability of the stochastic model, especially where direct experimental validation is constrained.

Efforts to explore the formation of domain structures in uniaxial ferroelectric PGO by simulations were initiated by Bak *et al.* [34] using the isotropic energy functional (1), although the focus was solely on a single domain nucleation, neglecting further domain structure formation. One potential challenge associated with the phase-field approach is the computational complexity of performing calculations in the Fourier space, which can be orders of magnitude faster than in the real space. Simulations in direct space were made with separately treating of the elastic field [102 – 104]. Mechanical conditions were obtained in each step by separate solving of elastic problem with use of finite element method. This avoids transformation in Fourier space and does not require periodicity of boundary conditions, but extremely increases the total calculation time. While the Fourier transform approach has been successfully implemented for ferroelectrics with cubic symmetry in the paraelectric phase (e.g., BTO, PTO, PZT, STO), adapting them for monoclinic TGS and triclinic PGO is considerably more intricate. Until the analytical groundwork addressing this task is completed, utilizing the phase-field method for uniaxial ferroelectrics remains confined to real space, demanding substantial computational resources even for minute objects spanning over several nanometers. A more advanced phase-field modelling of PGO by Tikhonov *et al*. [97] including the electrostriction and elasticity contributions allowed insight into exotic static domain structures but not an observation of the temporal evolution on a realistic time scale.

*Theory.* Considering the aforementioned challenges, the advancement of an analytical stochastic theory becomes imperative to effectively describe experimental observations by statistical characteristics and develop predictive tools. This framework facilitates the exploration of long-term dynamics and can be adapted to characterize various materials by adjusting the LGD functional [68, 105] for specific purposes. However, even when dealing with an isotropic problem, the derived equations necessitate numerical solutions [68 – 70]. Incorporating anisotropy into the model promises improved description of experiments conducted on TGS crystals, yet it introduces further significant complexities, raising concerns regarding the analytical solvability of equations. It was furthermore suggested by McCluskey et al. [93], that explanation of stable H-H and T-T domain configurations in uniaxial ferroelectrics requires rotation of polarization direction within these exotic domain walls. For incorporation of this reasonable idea in the stochastic theory the extension of the energy functional (1) to the full set of polarization components is necessary which is in the making.

Moreover, extending stochastic theory to encompass perovskite materials, which exhibit ferroelastic behavior, presents additional mathematical hurdles by introducing further stochastic variables. Inclusion of elasticity and electrostriction within the LGD functional poses challenges, compounded by the complexities associated with multi-stage polarization switching [106 – 109]. Addressing these mathematical intricacies will be crucial for the comprehensive modeling of perovskite materials and advancing our understanding of their behavior.

Further exploration of the processes governing temporal domain formation and ordering in ferroelectrics demands advancement across all fronts, encompassing analytical, experimental, and simulation methods. By enhancing analytical approaches such as the stochastic theory, we can refine our understanding of ferroelectric behavior and develop more accurate predictive models. Concurrently, experimental techniques should continue to evolve to enable precise observation and characterization of domain structures over realistic space and time scales under varied conditions. Additionally, simulation approaches need refinement to capture the intricate dynamics of domain evolution, particularly in materials with complex properties. By fostering progress in each of these directions, we can deepen our insights into nonlinear ferroelectric phenomena and pave the way for their future technological implementations.

**ACKNOWLEDGEMENTS**

O.Y.M. acknowledges the financial support by the Marie Sklodowska-Curie Actions for Ukraine (№ 1233427). This project has received funding through the MSCA4Ukraine project, which is funded by the European Union. Views and opinions expressed are however those of the author(s) only and do not necessarily reflect those of the European Union. Neither the European Union nor the MSCA4Ukraine Consortium as a whole nor any individual member institutions of the MSCA4Ukraine Consortium can be held responsible for them.

**DECLARATION OF INTERESTS**

The authors declare that they have no known competing financial interests or personal relationships that could have appeared to influence the work reported in this paper.

**DATA AVAILABILITY**

The data that support the findings of this study are available from the corresponding author upon reasonable request.

**REFERENCES**

1. J. Guo, W. Chen, H. Chen, Y. Zhao, F. Dong, W. Liu, and Y. Zhang, Recent progress in optical control of ferroelectric polarization, Adv. Optical Mater. 9 (2021) 2002146. https://doi.org/10.1002/adom.202002146

2. S.I. Shkuratov, C.S. Lynch, A review of ferroelectric materials for high power devices, Journal of Materiomics 8 (2022) 739. https://doi.org/10.1016/j.jmat.2022.04.002

3. H. Tan, A. Quintana, N. Dix, S. Estandia, J. Sort, F. Sanchez, I. Fina, Photovoltaic-driven dual optical writing and non-destructive voltage-less reading of polarization in ferroelectric $Hf_{0.5}Zr_{0.5}O_2$ for energy efficient memory devices, Nano Energy 123 (2024) 109384. https://doi.org/10.1016/j.nanoen.2024.109384

4. D. Wang, S. Hao, B. Dkhil, B. Tian, C. Duan, Ferroelecric materials for neuroinspired computing applications, Fundamental Research (2023). https://doi.org/10.1016/j.fmre.2023.04.013

5. M. Li, Z. Zou, Z. Xu, J. Zheng, Y. Li, R. Tao, Z. Fan, G. Zhou, X. Lu, J. Liu, Ferroelectric polarization and conductance filament coupling for large window and high-reliability resistive memory and energy-efficient synaptic devices, Journal of Materials Science & Technology 198 (2024) 36. https://doi.org/10.1016/j.jmst.2024.01.039

6. N. Sinha, S. Bhandari, H. Yadav, G. Ray, S. Godara, N. Tyagi, J. Dalal, S. Kumar, and B. Kumar, Performance of crystal violet doped triglycine sulfate single crystals for optical and communication applications, Cryst. Eng. Comm. 17 (2015) 5757. https://doi.org/10.1039/C5CE00703H

7. J.M. Sudhakaran, J. Philip, Triglycine sulphate and its deuterated analog in polyurethane matrix to thermal/infrared detection: A comparison, Journal of Applied Polymer Science 132 (2015) 42250. https://doi.org/10.1002/APP.42250

8. R. Ghane-Motlagh, and P. Woias, A pyroelectric thin film of oriented Triglycine sulfate nanocrystals for thermal energy harvesting, Smart Mater. Struct. 28 (2019) 104002. https://doi.org/10.1088/1361-665X/ab3910

9. Z.-D. He, W.-C. Li, J.-L. Yang, H.-K. Xu, G.-X. Lai, Y.-D. Che, W.-L. Zhu, X.-D. Yang, and X.-Y. Chen, Tuning ferroelectric photovoltaic performance in $R3c$-$CuNbO_3$ through compressive strain engineering: a first-principles study, RSC Adv. 13 (2023) 34475. https://doi.org/10.1039/d3ra07275d

10. J. Wang, Z. Peng, J. Wang, D. Wu, Z. Yang, X. Chao, Band gap tunning to enhance photovoltaic response in $NaNbO_3$-based bulk ferroelectrics, Scripta Materialia 221 (2022) 114976. https://doi.org/10.1016/j.scriptamat.2022.114976

11. X. Yang, X. Gao, S. Zhang, J. Zhao, X. Zhang, X. Song, C. Lu, Y. Li, L. Zhang, X. Hao, Anomalous photovoltaic effect in $Na_{0.5}Bi_{0.5}TiO_3$-based ferroelectric ceramics based on domain engineering, Journal of Materiomics (2023) https://doi.org/10.1016/j.jmat.2023.10.010

12. M. Davis, D. Damjanovic, D. Hayem, and N. Setter, Domain engineering of the transverse piezoelectric coefficient in perovskite ferroeletrics, J. Appl. Phys. 98 (2005) 014102. https://doi.org/10.1063/1.1929091

13. S. Wada, K. Yako, H. Kakemoto, T. Tsurumi, T. Kiguchi, Enhanced piezoelectric properties of barium titanate single crystals with different engineered-domain sizes, J. Appl. Phys. 98 (2005) 014109. https://doi.org/10.1063/1.1957130

14. C. Qiu, B. Wang, N. Zhang, S. Zhang, J. Liu, D. Walker, Y. Wang, H. Tian, T.R. Shrout, Z. Xu, L.-Q. Chen, F. Li, Transparent ferroelectric crystals with ultrahigh piezoelectricity, Nature 577 (2020) 350. https://doi.org/10.1038/s41586-019-1891-y

15. H.-C. Thong, Z. Li, J.-T. Lu, C.-B.-W. Li, Y.-X. Liu, Q. Sun, Z. Fu, Y. Wei, K. Wang, Domain engineering in bulk ferroelectric ceramics via mesoscopic chemical

inhomogeneity, Advanced Science 9 (2022) 2200998. https://doi.org/10.1002/advs.202200998

16. A.J. Klomp, R. Khachaturyan, T. Wallis, K. Albe, and A. Grunebohm, Thermal stability of nanoscale ferroelectric domains by molecular dynamics modeling, Phys. Rev. Materials 6 (2022) 104411. https://doi.org/10.1103/PhysRevMaterials.6.104411

17. D. Meier, and S.M. Selbach, Ferroelectric domain walls for nanotechnology, Nat. Rev. Mater. 7 (2022) 157. https://doi.org/10.1038/s41578-021-00375-z

18. J. Schultheiß, G. Picht, J. Wang, Y.A. Genenko, L.Q. Chen, J.E. Daniels, and J. Koruza, Ferroelectric polycrystals: Structural and microstructural levers for property-engineering via domain-wall dynamics, Progr. Mater. Sci. 136 (2023) 101101. https://doi.org/10.1016/j.pmatsci.2023.101101

19. S. Pandya, G.A. Velarde, R. Gao, A.S. Everhardt, J.D. Wilbur, R. Xu, J.T. Maher, J.C. Agar, C. Dames, and L.W. Martin, Understanding the role of ferroelastic domains on the pyroelectric and electrocaloric effects in ferroelectric thin films, Adv. Mater. 31 (2019) 1803312. https://doi.org/10.1002/adma.201803312

20. G.F. Nataf, M. Guennou, J.M. Gregg, D. Meier, J. Hlinka, E.K.H. Salje, and J. Kreisel, Domain-wall engineering and topological defects in ferroelectric and ferroelastic materials, Nat. Rev. Phys. 2 (2020) 634. https://doi.org/10.1038/s42254-020-0235-z

21. K. Matyjasek, Influence of temperature on the depolarization process in telluric acid ammonium phosphate crystal, J. Phys. D: Appl. Phys. 35 (2002) 3027. https://doi.org/10.1088/0022-3727/35/22/319

22. K. Matyjasek, Analysis of the switching process in triglycine sulphate crystals based on the Avrami model, J. Phys. D: Appl. Phys. 34 (2001) 2211. https://doi.org/10.1088/0022-3727/34/14/317

23. H. Okino, J. Sakamoto, and T. Yamamoto, Cooling-rate dependent domain structures of Pb(Mg1/3Nb2/3)O3−PbTiO3 single crystals observed by contact resonance piezoresponse force microscopy, Jap. J. Appl. Phys. 43 (2004) 6808. https://doi.org/10.1143/JJAP.43.6808

24. H. Okino, J. Sakamoto, and T. Yamamoto, Cooling-rate dependence of dielectric constant and domain structures in (1−x)Pb(Mg1/3Nb2/3)O3–xPbTiO3 single crystals, Jap. J. Appl. Phys. 44 (2005) 7160. https://doi.org/10.1143/JJAP.44.7160

25. D. S. Kim, C. I. Cheon, S. S. Lee, and J. S. Kim, Effect of cooling rate on phase transitions and ferroelectric properties in 0.75BiFeO3−0.25BaTiO3 ceramics, Appl. Phys. Lett. 109 (2016) 202902. https://doi.org/10.1063/1.4967742

26. H. Muramatsu, H. Nagata, and T. Takenaka, Quenching effects for piezoelectric properties on lead-free (Bi1/2Na1/2)TiO3 ceramics, Jpn. J. Appl Phys 55 (2016) 10TB07. https://doi.org/10.7567/JJAP.55.10TB07

27. K. V. Lalitha, J. Koruza, and J. Rodel, Propensity for spontaneous relaxor-ferroelectric transition in quenched (Na1/2Bi1/2)TiO3−BaTiO3 compositions, Appl. Phys. Lett. 113 (2018) 252902. https://doi.org/10.1063/1.5053989

28. J. Ma, Y. Zhu, Y. Tang, M. Han, Y. Wang, N. Zhang, M. Zou, Y. Feng, W. Geng, and X. Ma, Modulation of charged a1/a2 domains and piezoresponses of tensile strained PbTiO3 films by the cooling rate, RSC Adv. 9 (2019) 13981. https://doi.org/10.1039/c9ra02485a

29. Y. Ehara, D. Ichinose, M. Kodera, T. Shiraishi, T. Shimizu, T. Yamada, K. Nishida, and H. Funakubo, Influence of cooling rate on ferroelastic domain structure for epitaxial (100)/(001)-oriented Pb(Zr, Ti)O3 thin films under tensile strain, Jpn. J. Appl. Phys. 60 (2021) SFFB07. https://doi.org/10.35848/1347-4065/ac10f7

30. J. Liu, W. Chen, B. Wang, and Y. Zheng, Theoretical methods of domain structures in ultrathin ferroelectric films: A review, Materials 7 (2014) 6502. https://doi.org/10.3390/ma7096502

31. L.-Q. Chen, Phase-field method of phase transitions/domain structures in ferroelectric thin films: A review, J. Am. Ceram. Soc. 91 (2008) 1835. https://doi.org/10.1111/j.1551-2916.2008.02413.x

32. P.R. Potnis, N.-T. Tsou, and J.E. Huber, A review of domain modelling and domain imaging techniques in ferroelectric crystals, Materials 4 (2011) 417. https://doi.org/10.3390/ma4020417

33. P. Marton and J. Hlinka, Simulation of domain patterns in $BaTiO_3$, Phase Transit. 79 (2006) 467. https://doi.org/10.1080/01411590600892351

34. O. Bak, T.S. Holstad, Y. Tan, H. Lu, D.M. Evans, K.A. Hunnestad, B. Wang, J.P.V. McConville, P. Becker, L. Bohaty, I. Lukyanchuk, V.M. Vinokur, A.T.J. van Helvoort, J.M. Gregg, L.-Q. Chen, D. Meier, A. Gruverman, Observation of unconventional dynamics of domain walls n uniaxial ferroelectric lead germanate, Adv. Func. Mat. 30 (2020) 2000284. https://doi.org/10.1002/adfm.202000284

35. R. Indergand, A. Vidyasagar, N. Nadkarni, and D.M. Kochmann, A phase-field approach to studying the temperature-dependent ferroelectric response of bulk polycrystalline PZT, J. Mech. Phys. Solids 144 (2020) 104098. https://doi.org/10.1016/j.jmps.2020.104098

36. R. Indergand, X. Bruant, D.M. Kochmann, Domain pattern formation in tetragonal ferroelectric ceramics, Journal of the Mechanics and Physics of Solids 181 (2023) 105426. https://doi.org/10.1016/j.jmps.2023.105426

37. Y. Cao, A. Morozovska, and S.V. Kalinin, Pressure induced switching in ferroelectrics: Phase-filed modeling, electrochemistry, flexoelectric effect, and bulk vacancy dynamics, Phys. Rev. B 96 (2017) 184109. https://doi.org/10.1103/PhysRevB.96.184109

38. I.S. Vorotiahin, A.N. Morozovska, E.A. Eliseev, and Y.A. Genenko, Control of domain states in rhombohedral lead zirconate titanate films via strains and surface charges, Adv. Electron. Mater. 8 (2022) 2100386. https://doi.org/10.1002/aelm.202100386

39. A. Grunebohm, M. Marathe, R. Khachaturyan, R. Schiedung, D.C. Lupascu, , V.V. Schwartsman, Interplay of domain structure and phase transitions: theory , experiment and functionality, J. Phys.: Condens. Matter 34 (2021) 073002. https://doi.org/10.1088/1361-648X/ac3607

40. Y.-H. Shin, I. Grinberg, I.-W. Chen, A.M. Rappe, Nucleation and growth mechanism of ferroelectric domain-wall motion, Nature 449 (2007) 881. https://doi.org/10.1038/nature06165

41. S. Liu, I. Grinberg, A.M. Rappe, Intrinsic ferroelectric switching from first principles, Nature 534 (2016) 360. https://doi.org/10.1038/nature18286

42. N.A. Tikhomirova, S.A. Pikin, L.A. Shuvalov, L.I. Dontsova, E.S. Popov, A.V. Shilnikov, L.G. Bulatova, Visualization of static and the dynamics of domain structure in triglycine sulfate by liquid crystals, Ferroelectrics 29 (1980) 145. https://doi.org/10.1080/00150198008008470

43. N. Nakatani, Ferroelectric domain structure in TGS just below the Curie point after heat treatment, Jpn. J. Appl. Phys. 24 (1985) L528. https://doi.org/10.1143/JJAP.24.L528

44. N. Tomita, H. Orihara, and Y. Ishibashi, Ferroelectric domain pattern evolution in quenched triglycine sulphate, J. Phys. Soc. Jpn. 58 (1989) 1190. https://doi.org/10.1143/JPSJ.58.1190

45. H. Orihara, N. Tomita, and Y. Ishibashi, Pattern evolution of ferroelectric domain structure in TGS quenched below phase transition point, Ferroelectrics 95 (1989) 45. https://doi.org/10.1080/00150198908245176

46. V. Likodimos, M. Labardi, and M. Allegrini, Kinetics of ferroelectric domains investigated by scanning force microscopy, Phys. Rev. B 61 (2000) 14440. https://doi.org/10.1103/PhysRevB.61.14440

47. V. Likodimos, M. Labardi, X. K. Orlik, L. Pardi, M. Allegrini, S. Emonin, and O. Marti, Thermally activated ferroelectric domain growth due to random defects, Phys. Rev. B 63 (2001) 064104. https://doi.org/10.1103/PhysRevB.63.064104

48. V. Likodimos, M. Labardi, and M. Allegrini, Domain pattern formation and kinetics on ferroelectric surfaces under thermal cycling using scanning force microscopy, Phys. Rev. B 66 (2002) 024104. https://doi.org/10.1103/PhysRevB.66.024104

49. E.Z. Luo, Z. Xie, J.B. Xu, I.H. Wilson, and L.H. Zhao, *In situ* observation of the ferroelectric-paraelectric phase transition in a triglycine sulfate single crystal by variable-temperature electrostatic force microscopy, Phys. Rev. B 61 (2000) 203. https://doi.org/10.1103/PhysRevB.61.203

50. N. Nakatani, Observation of ferroelectric domain structure in TGS, Ferroelectrics 413 (2011) 238. https://doi.org/10.1080/00150193.2011.554269

51. S.N. Drozhdin and O.M. Golitsyna, Temperature and time behavior of the parameters of the domain structure of triglycine sulfate crystals near the phase transition, Phys. Solid State 54 (2012) 905. https://doi.org/10.1134/S1063783412050071

52. G.I. Ovchinnikova, N.V. Belugina, R.V. Gainutdinov, E.S. Ivanova, V.V. Grebenev, A.K. Lashkova, and A.L. Tolstikhina, Temperature dynamics of triglycine sulfate domain structure according to atomic force microscopy and dielectric spectroscopy data, Phys. Solid State 58 (2016) 2244. https://doi.org/10.1134/S1063783416110305

53. L. Wehmeier, T. Kämpfe, A. Haußmann, and L. M. Eng, In situ 3D observation of the domain wall dynamics in a triglycine sulfate single crystal upon ferroelectric phase

transition, Phys. Status Solidi - Rapid Res. Lett. 11 (2017) 1700267. https://doi.org/10.1002/pssr.201700267

54. O.M. Golitsyna, S.N. Drozhdin, V.O. Chulakova, and M.N. Grechkina, Evolution of the domain structure of triglycine sulphate single crystal in the vicinity of phase transition, Ferroelectrics 506 (2017) 127. https://doi.org/10.1080/00150193.2017.1282286

55. O.M. Golitsyna and S.N. Drozhdin, Formation of a quasi-equilibrium domain structure of crystals of the TGS group near $T_C$, Condens. Matter Interphases 23 (2021) 507. https://doi.org/10.17308/kcmf.2021.23/3669

56. A.P. Turygin, M.S. Kosobokov, O.M. Golitsyna, S.N. Drozhdin, and V. Y. Shur, Unusual domain growth during local switching in triglycine sulfate crystals, Appl. Phys. Lett. 119 (2021) 262902. https://doi.org/10.1063/5.0077685

57. T. Nattermann, Static and dynamic critical behaviour of uniaxial ferroelectrics and the phase transition in TGS, Phys. Stat. Solidi 85 (1978) 291. https://doi.org/10.1002/pssb.2220850132

58. Y. Ishibashi and Y. Takagi, Note on ferroelectric domain switching, J. Phys. Soc. Jpn. 31 (1971) 506. https://doi.org/10.1143/JPSJ.31.506

59. A. N. Kolomgorov, On the statistical theory of crystallization of metals Izv. Akad. Nauk SSSR, Ser. Mat. 1 (1937) 355.

60. S. Hashimoto, H. Orihara and Y. Ishibashi, Study on D-E hysteresis loop of TGS based on the Avrami-type model, J. Phys. Soc. Jpn. 63 (1994) 1601. https://doi.org/10.1143/JPSJ.63.1601

61. F. Kalkum, H. A. Eggert, T. Jungk and K. Buse, A stochastic model for periodic domain structuring in ferroelectric crystals, J. Appl. Phys. 102 (2007) 014104. https://doi.org/10.1063/1.2752545

62. M. Rao and A. Chakrabarti, Kinetics of Domain Growth in a Random-Field Model in Three Dimensions, Phys. Rev. Lett. 71 (1993) 3501. https://doi.org/10.1103/PhysRevLett.71.3501

63. L. I. Stefanovich, Formation and growth dynamics of domains under phase transitions in an external field, Low Temp. Phys. 24 (1998) 643. https://doi.org/10.1063/1.593652

64. K. Gumennyk, L. Stefanovich, and E. Feldman, Kinetics of coupled ordering and segregation in antiphase domains, Phys. Status Solidi B 246 (2009) 56. https://doi.org/10.1002/pssb.200844277

65. E.P. Feldman, L.I. Stefanovich, and Y.V. Terekhova, Influence of adsorption or desorption an surface diffusion on the formation kinetics of open half-monolayer coverage, Phys. Rev. B 89 (2014) 062406. https://doi.org/10.1103/PhysRevE.89.062406

66. B. M. Darinskii, A. P. Lazarev, and A. S. Sigov, JETP 87 (1998) 1221. https://doi.org/10.1134/1.558616

67. A.J. Bray, Theory of phase-ordering kinetics, Advances in Physics 43 (1994) 357. https://doi.org/10.1080/00018739400101505

68. O.Y. Mazur, L.I. Stefanovich, and Y.A. Genenko, Stochastic theory of ferroelectric domain structure formation dominated by quenched disorder, Phys. Rev. B 107 (2023) 144109. https://doi.org/10.1103/PhysRevB.107.144109

69. Y.A. Genenko, O.Y. Mazur, and L.I. Stefanovich, Statistical correlations of random polarization and electric depolarization fields in ferroelectrics, Phys. Rev. B 108 (2023) 134101. https://doi.org/10.1103/PhysRevB.108.134101

70. O.Y. Mazur, L.I. Stefanovich, and Y.A. Genenko, Universal kinetics of stochastic formation of polarization domain structures in a uniaxial single-crystal ferroelectric, Phys. Rev. B 109 (2024) 104117. https://doi.org/10.1103/PhysRevB.109.104117

71. A. K. Tagantsev, L. E. Cross, J. Fousek, Domains in Ferroic Crystals and Thin Films, Springer, New Jork, 2010.

72. O.Yu. Mazur, L.I. Stefanovich, V.M. Yurchenko, Influence of Quenching Conditions on the Kinetics of Formation of a Domain Structure of Ferroelectrics, Physics of the Solid State 57 (2015) 576–585. https://doi.org/10.1134/S1063783415030142

73. R. Landauer, Electrostatic considerations in $BaTiO_3$ domain formation during polarization reversal, J. Appl. Phys. 28 (1957) 227. https://doi.org/10.1063/1.1722712

74. R.C. Miller, G. Weinreich, Mechanism for the sidewise motion of 180º domain walls in barium titanate, Phys. Rev. 117 (1960) 1460. https://doi.org/10.1103/PhysRev.117.1460

75. G. Gerra, A.K. Tagantsev, N. Setter, Surface-stimulated nucleation of reverse domains in ferroelectrics, Phys. Rev. Lett. 94 (2005) 107602. https://doi.org/10.1103/PhysRevLett.94.107602

76. H. Vogt and D. Weinmann, Optical second harmonic generation during ferroelectric polarization reversal, Phys. Stat. Sol. 14 (1972) 501. https://doi.org/10.1002/pssa.2210140215

77. V.Y. Shur, V.A. Sikhova, D.V. Pelegov, A.V. Ievlev, and L.I. Ivleva, Formation of Nanodomain Ensembles during Polarization reversal in $Sr_{0.61}Ba_{0.39}Nb_2O_6$: Ce single crystals, Physics of the Solid State 53 (2011) 2311. https://doi.org/10.1134/S106378341111028X

78. A.R. Akhmatkhanov, M.A. Chuvakova, M.S. Nebogatikov, Y.V. Shaydurov, and V.Y. Shur, Domain splitting in lithium niobate with surface dielectric layer, Ferroelectrics 559 (2020) 8. https://doi.org/10.1080/00150193.2020.1721999

79. K. Li, E. Sun, X. Qi, B. Yang, J. Liu, W. Cao, Dielectric relaxation and local domain structures of ferroelectric PIMNT and PMNT single crystals, J. Am. Ceram. Soc. (2019). https://doi.org/10.1111/jace.16849

80. X. Qi, K. Li, L. Bian, E. Sun, L. Zheng, and R. Zhang, Domain structure and dielectric diffusion-relaxation characteristics of ternary Pb(In$_{1/2}$Nb$_{1/2}$)O$_3$-Pb(Mg$_{1/3}$Nb$_{2/3}$)O$_3$-PbTiO$_3$ ceramics, Journal of Advanced Dielectrics 12 (2022) 2241002. https://doi.org/10.1142/S2010135X22410028

81. V.V. Shvartsman, A.L. Kholkin, A. Orlova, D. Kiselev, A.A. Bogomolov, and A. Sternberg, Polar nanodomains and local ferroelectric phenomena in relaxor lead lanthanum zirconate titanate ceramics, Appl. Phys. Lett. 86 (2005) 202907. http://dx.doi.org/10.1063/1.1923756

82. V.V. Shvartsman, A.L. Kholkin, Evolution of nanodomains in 0.9PbMg$_{1/3}$Nb$_{2/3}$O$_3$-0.1PbTiO$_3$, J. Appl. Phys. 101 (2007) 064108. http://dx.doi.org/10.1063/1.2713084

83. V.V. Shvartsman, J. Dec, T. Lukasiewicz, A.L. Kholkin, and W. Kleemann, Evolution of the polar structure in relaxor ferroelectrics close to the curie temperature studied by piezoresponse force microscopy, Ferroelectrics 373 (2008) 77. https://doi.org/10.1080/00150190802408739

84. N. Balke, I. Bdikin, S.V. Kalinin, and A.L. Kholkin, Electromechanical imaging and spectroscopy of ferroelectric and piezoelectric materials: state of the art and prospects for the future, J. Am. Ceram. Soc. 92 (2009) 1629. https://doi.org/10.1111/j.1551-2916.2009.03240.x

85. P. Bintachitt, S. Trolier-McKinstry, K. Seal, S. Jesse and S. V. Kalinin, Switching spectroscopy piezoresponse force microscopy of polycrystalline capacitor structures, Appl. Phys. Lett. 94 (2009) 042906. https://doi.org/10.1063/1.3070543

86. K. Seal, S. Jesse, M. P. Nikiforov, S. V. Kalinin, I. Fujii, P. Bintachitt and S. Trolier-McKinstry, Spatially resolved spectroscopic mapping of polarization reversal in polycrystalline ferroelectric films: crossing the resolution barrier, Phys. Rev. Lett. 103 (2009) 057601. https://doi.org/10.1103/PhysRevLett.103.057601

87. A. L. Tolstikhina, N. V. Belugina and R. V. Gainutdinov, Application of the Fourier transforms for analysis of the domain structure images of uniaxial ferroelectric, Crystallogr. Rep. 62 (2017) 464. https://doi.org/10.1134/S1063774517030221

88. C. Weymann, S. Cherifi-Hertel, C. Lichtensteiger, I. Gaponenko, K.D. Dorkenoo, A.B. Naden and P. Paruch, Non-Ising domain walls in *c*-phase ferroelectric lead titanate thin films, Phys. Rev. B 106 (2022) L241404. https://doi.org/10.1103/PhysRevB.106.L241404

89. C. Lichtensteiger, M. Hadjimichael, E. Zatterin, C.-P. Su, I. Gaponenko, L. Tovaglieri, P. Paruch, A. Gloter, and J.-M. Triscone, Mapping the complex evolution of ferroelastic/ferroelectric domain patterns in epitaxially strained $PbTiO_3$ heterostructures, APL Mater. 11 (2023) 061126. https://doi.org/10.1063/5.0154161

90. H. Vogt, Study of structural phase transitions by techniques of nonlinear optics, Appl. Phys. 5 (1974) 85. https://doi.org/10.1007/BF00928219

91. G. Dolino, Effects of domain shapes on second-harmonic scattering in triglycine sulfate, 6 (1972) 4025. https://doi.org/10.1103/PhysRevB.6.4025

92. O. Nesterov, S. Matzen, C. Magen, A.H.G. Vlooswijk, G. Catalan, and B. Noheda, Thickness scaling of ferroelastic domains in $PbTiO_3$ films on $DyScO_3$, Appl. Phys. Lett. 103 (2013) 142901. https://doi.org/10.1063/1.4823536

93. C.J. McCluskey, A. Kumar, A. Gruverman, I. Luk'yanchuk, and J.M. Gregg, Domain wall saddle point morphology in ferroelectric triglycine sulfate, Appl. Phys. Lett. 122 (2023) 222902. https://doi.org/10.1063/5.0152518

94. D. Wolf, A. Lubk, P. Prete, N. Lovergine, and H. Lichte, 3D mapping of nanoscale electric potentials in semiconductor structures using electron-holographic tomography, Phys. D: Appl. Phys. 49 (2016) 364004. https://doi.org/10.1088/0022-3727/49/36/364004

95. R. Khachaturyan and Y.A. Genenko, Correlated polarization-switching kinetics in bulk polycrystalline ferroelectrics. II. Impact of crystalline phase symmetries, Phys. Rev. B 98 (2018) 134106. https://doi.org/10.1103/PhysRevB.98.134106

96. Y. Tikhonov, J.R. Maguire, C.J. McCluskey, J.P.V. McConville, A. Kumar, H. Lu, D. Meier, A. Razumnaya, J.M. Gregg, A. Gruverman, V.M. Vinokur, and I. Luk'yanchuk, Polarization topology at the nominally charged domain walls in uniaxial ferroelectrics, Advanced Materials 34 (2022) 2203028. https://doi.org/10.1002/adma.202203028

97. M.L. Mulvmill, K. Uchino, Z. Li, and W. Cao, *In-situ* observation of the domain configurations during the phase transitions in barium titanate, Philosophical Magazine B 74 (1996) 25. https://doi.org/10.1080/01418639608240325

98. E.A. Little, Dynamic behavior of domain walls in barium titanate, Phys. Rev. 98, 978 (1955). https://doi.org/10.1103/PhysRev.98.978

99. M. McQuarrie, Role of domain processes in polycrystalline barium titanate, J. Am. Ceram. Soc. 39 (1956) 54. https://doi.org/10.1111/j.1151-2916.1956.tb15623.x

100. P. Marton, I. Rychetsky, and J. Hlinka, Domain walls of ferroelectric $BaTiO_3$ within the Ginzburg-Landau-Devonshire phenomenological model, Phys. Rev. B 81 (2010) 114125. https://doi.org/10.1103/PhysRevB.81.144125

101. D.A. Kiselev, T.S. Ilina, M.D. Malinkovich, O.N. Sergeeva, N.N. Bolshakova, E.M. Semenova, and Y.V. Kuznetsova, Specific features of the domain structure of $BaTiO_3$ crystals during thermal heating and cooling, Physics of the Solid State 60 (2018) 738. https://doi.org/10.1134/S1063783418040157

102. S. Nambu and D.A. Sagala: Domain formation and elastic long-range interaction in ferroelectric perovskites, Phys. Rev.B 50 (1994) 5838. https://doi.org/10.1103/PhysRevB.50.5838

103. W. Zhang and K. Bhattachatya: A computational model of ferroelectric domains. Part I: model formulation and domain switching, Acta Mater. 53 (2005) 185. https://doi.org/10.1016/j.actamat.2004.09.016

104. W. Zhang and K. Bhattachatya: A computational model of ferroelectric domains. Part II: grain boundaries and defect pinning, Acta Mater. 53 (2005) 199. https://doi.org/10.1016/j.actamat.2004.09.015

105. O. Mazur, K. Tozaki, Y. Yoshimura, and L. Stefanovich, Influence of pressure on the kinetics of ferroelectric phase transition in BaTiO3, Physica A 599 (2022) 127436. https://doi.org/10.1016/j.physa.2022.127436

106. Y.A. Genenko, R. Khachaturyan, J. Schultheiß, A. Ossipov, J.E. Daniels, and J. Koruza, Stochastic multistep polarization switching in ferroelectrics. Phys. Rev. B 97 (2018) 144101. https://doi.org/10.1103/PhysRevB.97.144101

107. J. Schultheiß, L. Liu, H. Kungl, M. Weber, L. Kodumudi Venkataraman, S. Checchia, D. Damjanovic, J.E. Daniels, and J. Koruza, Revealing the sequence of switching mechanisms in polycrystalline ferroelectric/ferroelastic materials, Acta Mater. 157 (2018) 355. https://doi.org/10.1016/j.actamat.2018.07.018

108. Y.A. Genenko, R. Khachaturyan, I.S. Vorotiahin, J. Schultheiß, J.E. Daniels, A. Grünebohm, and J. Koruza, Multistep stochastic mechanism of polarization reversal in rhombohedral ferroelectrics. Phys. Rev. B 102 (2020) 064107. https://doi.org/10.1103/PhysRevB.102.064107

109. Y.A. Genenko, M.-H. Zhang, I.S. Vorotiahin, R. Khachaturyan, Y.-X. Liu, J.-W. Li, K. Wang, and J. Koruza, Multistep stochastic mechanism of polarization reversal in orthorhombic ferroelectrics. Phys. Rev. B 104 (2021) 184106. https://doi.org/10.1103/PhysRevB.104.184106